\documentclass[twocolumn]{aastex62}
\usepackage{amsmath}
\usepackage{graphicx}
\usepackage{amsmath}

\begin{document}

\title{ELT Imaging of MWC 297 from the 23-m LBTI: Complex Disk Structure and a Companion Candidate}

\author{S. Sallum}
\altaffiliation{Department of Physics and Astronomy, 4129 Frederick Reines Hall,\\ University of California, Irvine, CA, 92697-4575, USA}

\author{J.A. Eisner}
\altaffiliation{Astronomy Department, University of Arizona,\\ 933 N. Cherry Ave., Tucson, AZ 85721, USA}

\author{J.M. Stone}
\altaffiliation{Astronomy Department, University of Arizona,\\ 933 N. Cherry Ave., Tucson, AZ 85721, USA}
\altaffiliation{Naval Research Laboratory, Remote Sensing Division,\\ 4555 Overlook Ave SW, Washington, DC 20375, USA}
\altaffiliation{Hubble Fellow}

\author{J. Dietrich}
\altaffiliation{Astronomy Department, University of Arizona,\\ 933 N. Cherry Ave., Tucson, AZ 85721, USA}

\author{P. Hinz}
\altaffiliation{Astronomy Department, University of California Santa Cruz,\\ 1156 High St., Santa Cruz, CA 95064, USA}

\author{E. Spalding}
\altaffiliation{Astronomy Department, University of Arizona,\\ 933 N. Cherry Ave., Tucson, AZ 85721, USA}

\correspondingauthor{Steph Sallum}
\email{ssallum@uci.edu}

\begin{abstract}

Herbig Ae / Be stars represent the early outcomes of star formation and the initial stages of planet formation at intermediate stellar masses.
Understanding both of these processes requires detailed characterization of their disk structures and companion frequencies.
We present new 3.7 $\mu$m imaging of the Herbig Be star MWC 297 from non-redundant masking observations on the phase-controlled, 23-m Large Binocular Telescope Interferometer.
The images reveal complex disk structure on the scales of several au, as well as a companion candidate.
We discuss physical interpretations for these features, and demonstrate that the imaging results are independent of choices such as priors, regularization hyperparameters, and error bar estimates.
With an angular resolution of $\sim17$ mas, these data provide the first robust ELT-resolution view of a distant young star.\\
\end{abstract}

\section{Introduction}
Herbig Ae/Be stars are young, intermediate-mass stars hosting protoplanetary disks \citep[e.g.][]{1960ApJS....4..337H,1992ApJ...397..613H}, often thought to represent a transition in formation mechanism between high- and low-mass stars \citep[e.g][]{2005MNRAS.359.1049V}. 
Observing their disks in detail and placing constraints on their companion occurrences presents an opportunity to study the physics of star formation, and to probe the initial stages of planet formation around massive stars.
Interferometric studies of these objects have revealed extended millimeter emission \citep[e.g.][]{2003A&A...398..565P,2009A&A...497..117A}, compact infrared circumstellar disks \citep[e.g.][]{2001ApJ...546..358M,2004ApJ...613.1049E}, and winds and outflows \citep[e.g.][]{2007A&A...464...43M}.
Furthermore, a variety of surveys using visual and spectroscopic techniques have revealed a high companion frequency \citep[$30-75\%$; e.g.][]{1997A&A...318..472L,2001IAUS..200..155B,1999A&AS..136..429C,2006MNRAS.367..737B}, with a possibly higher frequency for Be stars than for Ae stars \citep[e.g.][]{2006MNRAS.367..737B}.
Herbig Ae/Be stars are thus also a unique laboratory for disk-companion interactions. 

\subsection{MWC 297}
Here we present new, spatially-resolved observations of the Herbig Be star MWC 297, which is located in the Aquila Rift.
Its mass and age are estimated to be $\sim17~\mathrm{M_\odot}$ and $\sim2.8\times10^4$ yr, respectively \citep[e.g.][]{2018A&A...620A.128V}. 
Its spectral energy distribution classifies it as a Meeus Group I object, indicating the presence of a circumstellar disk that contributes significantly to the infrared luminosity \citep[e.g][]{2001A&A...365..476M}.
Previous studies have yielded a variety of distance and extinction estimates to MWC 297, constrained its rotational velocity, and studied its circumstellar environment at a range of wavelengths.
The following three subsections detail the various distance estimates (\ref{sec:intro:dist}), disk characterization efforts (\ref{sec:intro:disk}), and constraints on complex asymmetric structures (e.g. companions and/or winds) in the context of MWC 297's stellar properties (\ref{sec:intro:comp}).

\subsection{The Distance to MWC 297}\label{sec:intro:dist}
Estimates of MWC 297's distance have ranged from 250 pc to $>$450 pc. 
Its distance was initially constrained to be $<600$ pc, based on a B0-or-later spectral type and and an extinction of $A_v\sim6.7$ mag derived from HI line flux ratios \citep{1977ApJ...218..170T,1988Ap.....28..313B}.
CO kinematic observations decreased the distance estimate to $\sim450$ pc \citep{1984ApJ...282..631C}.
Followup optical spectroscopy constrained the spectral type, $A_v$, and distance simultaneously, yielding B1.5V (within 0.5 subtypes), $A_v \sim 7.8$ mag, and $250\pm50$ pc \citep{1997MNRAS.286..538D}.\footnote{When calculating the distance to MWC 297, \citet{1997MNRAS.286..538D} assumed $A_v = 8.0$ mag for consistency with previous literature, rather than their own estimate of $A_v = 7.8$ mag.} 
In contrast, \emph{Gaia} recently estimated the distance to MWC 297 to be $375\pm20$ pc \citep{2018A&A...620A.128V}.

Of the distance measures above, the CO kinematic distance is highly uncertain, since this method is unreliable within 1 kpc \citep{1984ApJ...282..631C}.
The remaining discrepancy lies between the photometric distance of $250\pm50$ pc and the \emph{Gaia} distance of $375\pm20$ pc. 
The photometric distance relies on estimates of MWC 297's V band magnitude and extinction, which are both uncertain. 
MWC 297 has been shown to be variable in V band, with apparent magnitudes ranging from $\sim12.0-12.5$ \citep{1988Ap.....28..313B}. 
The majority of extinction estimates range from $A_v\sim7.8-8.3$ mag \citep{1984ApJ...286..609M,1992ApJ...397..613H,1997MNRAS.286..538D}, with the exception of $A_v\sim6.7$ mag from \citet{1977ApJ...218..170T}. 
However, the H I lines in  \citet{1977ApJ...218..170T} were not observed simultaneously, and MWC 297 is known to be variable in Hydrogen lines \citep[e.g.][]{2015MNRAS.447..202E}, making this estimate less reliable.

Given the uncertainties for V magnitude and extinction, we explore whether the \emph{Gaia} distance is consistent with a B1.5($\pm$0.5)V star.
Following the same procedure as in \citet{1997MNRAS.286..538D}, MWC 297's range in apparent V magnitude of $12.0-12.5$ and its extinction estimates of $A_v = 7.8-8.3$ yield $\mathrm{M_V} =$  -3.0 to -4.0 mag for a distance of 375 pc, consistent with expected $\mathrm{M_V}$ values for B0V - B1.5V spectral types \citep[e.g.][]{1981Ap&SS..80..353S}.
This calculation is in agreement with more recent 0.3-1.3 $\mu$m spectral energy distribution modeling of MWC 297, which derived $A_v\sim7.7$ for a Kurucz B1.5V model spectrum \citep{1991ppag.conf...27K} assuming a distance of 375 pc \citep{2020ApJ...890L...8U}.

The remaining argument for a shorter distance of $250\pm50$ pc is that as a young star MWC 297 is likely associated with the Aquila Rift.
Photometry suggests that the closest edge of the cloud complex lies at $225\pm55$ pc, and that the complex has a thickness of $\sim 80$ pc \citep{2003A&A...405..585S}.
The far edge of the cloud would thus lie between 250 pc and 360 pc, making a 375 pc distance to MWC 297 an outlier at the $1\sigma$ level.
However, VLBA parallax measurements to young stars in the vicinity of MWC 297 estimate their distances to be much larger \citep[$>400$ pc e.g.][]{2017ApJ...834..143O}.
A 375 pc distance to MWC 297 is thus within the range of distance estimates to the Aquila Rift.
For this reason and due to its agreement with MWC 297's extinction and spectral type estimates, we adopt the \emph{Gaia} distance of 375 pc for this work, and adjust any spatial measurements from previous studies to this distance.

\subsection{MWC 297's Circumstellar Disk}\label{sec:intro:disk}
MWC 297 hosts a circumstellar disk that has been well studied in the radio.
An early 5 GHz (6 cm) map revealed extended structure on $\sim200-300$ mas scales, and a north-south elongation \citep{1997MNRAS.286..538D}.
Given the large $v \sin{i}$ measured from He I lines ($\sim350$ km/s), this study suggested that the north-south elongation may trace an edge-on circumstellar disk \citep{1997MNRAS.286..538D}.
Followup observations at 1.3 mm and 2.7 mm did not show the same complex structure \citep{2009A&A...497..117A}, but they had poorer angular resolution (beam sizes of $1.1"\times1.4"$ and $1.4"\times0.9"$, respectively, compared to $0.14"\times0.11"$).
A joint fit to these data and MWC 297's SED resulted in a best-fit outer disk radius of $\sim43$ au at 375 pc, similar to the extent of the 6 cm emission. 
The observations were well explained by either an $i\sim5^\circ$ disk with an inner rim, or an $i\sim80^\circ$ disk without an inner rim.
The 1.3mm/2.7mm spectral slope suggested that a ring of large ($\sim1$ cm) grains may also exist at radii of $200-300$ au \citep{2009A&A...497..117A}.

MWC 297 has been observed in the infrared with both long-baseline (Michelson) and Fizeau interferometry.
Studies utilizing VLTI/MIDI and VLTI/AMBER \citep{2008A&A...485..209A,2007A&A...464...43M,2011A&A...527A.103W,2017A&A...607A..17H}, VLTI/PIONIER \citep{2017A&A...599A..85L,2020A&A...636A.116K}, IOTA \citep{2001ApJ...546..358M,2006ApJ...647..444M}, and PTI \citep{2004ApJ...613.1049E} have characterized the structure of the circumstellar disk at H band ($1.0-1.6~\mu$m), K band ($2.0-2.5~\mu$m), and N band ($8-13~\mu$m).
The near-IR measurements found best-fit Gaussian FWHMs of $\sim2-5$ mas, and a symmetric geometry consistent with a relatively face-on disk ($i\sim15-38^\circ$).
Fits to N band MIDI observations found evidence for more extended structure \citep[FWHM $\sim$ 40 mas; e.g][]{2008A&A...485..209A}, consistent with a best-fit FWHM of $\sim60$ mas from $10.7\mu$m segment-tilting Fizeau interferometry at Keck \citep{2009ApJ...700..491M}.

\subsection{Rotational Velocity, Complex Disk Structure, and Companion Scenarios} \label{sec:intro:comp}

MWC 297's low inclination estimates are at odds with a spectrosopic $v\sin{i}$ measurement of $\mathbf{350\pm50~\mathrm{km~s^{-1}}}$ \citep{1997MNRAS.286..538D}.
For this $v\sin{i}$, inclinations of $15-38^\circ$ correspond to rotational velocities $\gtrsim550~\mathrm{km~s^{-1}}$.
However, recent estimates of MWC 297's mass and radius, $\sim17~\mathrm{M_\odot}$ and 9.7 $R_\odot$, respectively \citep{2018A&A...620A.128V,2020ApJ...890L...8U} suggest a breakup velocity of $\sim480\mathrm{~km~s^{-1}}$.
If the stellar $v\sin{i}$ is indeed $350~\mathrm{km~s^{-1}}$, then the stellar inclination must be $\gtrsim50^\circ$ in order for the rotational velocity to be below breakup.

For a star with a wind, some spectral features could be contaminated by wind material or formed in a location where outflow kinematics dominate \citep[e.g][]{2011A&A...527A.103W}, which could lead to an overestimated stellar $v\sin{i}$.
MWC 297's $v\sin{i}$ was measured in two ways: (1) by broadening standard star spectral lines and comparing them to MWC 297, and (2) computing the widths of He I singlets at 4009 $\mathrm{\AA}$ and 4144 $\mathrm{\AA}$ \citep{1997MNRAS.286..538D}.
While MWC 297 is known to have a wind \citep{2007A&A...464...43M,2011A&A...527A.103W}, the He I singlet $v\sin{i}$ ($\sim320~\mathrm{km~s^{-1}}$) is unlikely to be contaminated by circumstellar material given the high line excitations.

If the $v\sin{i}$ measurement errors were underestimated, then the stellar rotation rate could be consistent with a $\lesssim38^\circ$ inclination. 
If the true $v\sin{i}$ were $\sim290~\mathrm{km~s^{-1}}$ then the stellar rotation would be below breakup for the most recent inclination estimate of $38^\circ$ \citep{2020A&A...636A.116K}. 
This $v\sin{i}$ would correspond to a one pixel change in the He I line widths measured in \citet{1997MNRAS.286..538D}, using the Intermediate-dispersion Spectrograph and Imaging System on the William Herschel Telescope.
It would also only be $10~\mathrm{km~s^{-1}}$ lower than the nominal 1$\sigma$ bounds on the measurement ($300-400~\mathrm{km~s^{-1}}$).

It is also possible that MWC 297's inclination is underestimated. 
The inclination constraints are based on long-baseline visibilities that indicate a symmetric brightness distribution on $\sim$ few milliarcsecond scales \citep[e.g.][]{2007A&A...464...43M,2011A&A...527A.103W,2017A&A...607A..17H}.
Recent VLTI data including phase information reveals asymmetric emission that could be explained by disk inclination effects for $i\sim38^\circ$ \citep{2020A&A...636A.116K}.
However, it is also possible that complex circumstellar structure could cause a brightness distribution that appears relatively symmetric despite a higher stellar inclination. 
For example, a disk wind could produce this effect, since dust can become entrained in the wind \citep[e.g.][]{2012ApJ...758..100B}.
This would cause the brightness distribution to appear more spherical on the sky even at a high disk inclination, which could lead to an underestimated disk inclination in geometric fitting.

An outer companion could cause a symmetric brightness distribution for a high stellar inclination by inducing a disk warp.
This could cause the disk inclination to differ from the stellar inclination for at least some spatial separations.
This scenario could be tested by searching for shadows or asymmetries in scattered light on the scales of a few au, and searching for companions that may be perturbing the disk on those scales.
There are no reports of disk shadowing or warps in the innermost several au around MWC 297 in the literature.

VLT/SPHERE recently discovered a sub-stellar companion ($\mathrm{M}\sim0.1-0.5~\mathrm{M_\odot}$) around MWC 297 with a separation of $\sim245$ au \citep{2020ApJ...890L...8U}. 
However, given its mass and separation, following the first order perturbation theory in \citet{1993A&A...274..291T}, this companion is not massive enough to cause large warps in the inner $\sim$ few to tens of au. 
A companion-induced disk warp would require a companion with a smaller separation and/or higher mass.
The VLT/SPHERE observations did not detect any other companions down to their coronagraph inner working angle 150 mas (56 au at 375 pc).
This inner working angle also makes these data poorly suited for searching for inner disk warps and shadows in the innermost few tens of au.

\subsection{Outline of This Paper}

Here we present 3.7 $\mu$m ELT-resolution imaging of the Herbig Be star MWC 297, from the co-phased Large Binocular Telescope Interferometer \citep[LBTI;][]{2008SPIE.7013E..28H,2014SPIE.9148E..03B}.
With a 23-meter effective aperture, LBTI can robustly image MWC 297's circumstellar emission down to spatial scales of several au. 
We place new constraints on the circumstellar disk geometry and detect a companion candidate with a separation of $\sim22$ au.

In Sections \ref{sec:exp}, \ref{sec:obs} and \ref{sec:red} we describe the experimental design, observations, and data reduction, respectively. 
Sections \ref{sec:ana} and \ref{sec:results} discuss the image reconstruction process and present its results. 
We discuss MWC 297's morphology in Section \ref{sec:disc} and summarize the main conclusions in Section \ref{sec:conc}.
Appendices \ref{app:red}, \ref{app:modeling}, and \ref{app:imrecon} provide additional details regarding data reduction, geometric modeling, and image reconstruction tests for exploring image fidelity.

\section{Experimental Design}\label{sec:exp}
The technique of non-redundant masking \citep[NRM;][]{2000PASP..112..555T} has been demonstrated to provide useful constraints on Herbig Ae/Be stellar environments \citep[e.g.][]{2001Natur.409.1012T,2017ApJ...844...22S}.
NRM uses a pupil-plane mask to transform a conventional telescope into an interferometric array, making the images on the detector the interference fringes formed by the mask.
These images are Fourier transformed to calculate complex visibilities, which have both amplitude and phase.
Since the baselines in the array are non-redundant (no repeated lengths and position angles), information from each baseline has a unique spatial frequency.
From the complex visibilities we calculate squared visibilities, the powers associated with the mask baselines.
We also calculate closure phases, sums of phases around baselines forming a triangle.
Non-redundancy means that closure phases eliminate residual instrumental and atmospheric phase errors to first order, making them particularly powerful for close-in companion detection.

NRM provides moderate contrast down to separations of $\sim0.5~\lambda/D$ \citep[e.g.][]{2008ApJ...678L..59I,2019JATIS...5a8001S}, a factor of a few to several boost in resolution compared to high performance coronagraphy \citep[e.g.][]{2014ApJ...780..171G,2017AJ....154...73R}.
Due to its sparse Fourier coverage, model fitting and/or image reconstruction (often used in conjunction) are required to constrain the source brightness distribution.
Simulations have shown that with adequate $(u,v)$ sampling and sky rotation coverage, robust, model-independent images can be reconstructed from NRM observations \citep[e.g.][]{2017ApJS..233....9S}.

The NRM observations presented here were taken with operational co-phasing of the two 8.4-m LBT mirrors. 
With co-phasing, the wavefront across each mirror is flattened using adaptive optics and the differential piston between the two mirrors is controlled using a cryogenic pathlength corrector. 
These observations used the flexible pyramid wavefront sensors at 600-900 nm to correct 153 Zernike modes at 990 Hz.
Differential piston, tip and tilt were sensed at Ks-band using the phasecam fringe tracker run at 1 kHz. 
In this mode the LBT can be thought of as a segmented telescope with two 8-meter mirror segments, providing $\sim23$-m resolution in one direction, and $\sim8$-m resolution in the perpendicular direction.
While previous LBTI NRM observations have utilized inter-aperture baselines in ``lucky" imaging mode \citep{2017ApJ...844...22S}, these are the first data with operational co-phasing.

Applied on an Extremely Large Telescope (ELT; D $\gtrsim$ 25 m), 3.7 $\mu$m NRM observations resolve $\sim15$ mas angular scales, accessing the inner few to several au around distant ($\gtrsim500$ pc) young stars.
The $\sim1-23$ meter baselines offered by the co-phased LBTI resolve angular separations down to $\sim17$ mas at L$'$.
The dense ($u,v$) coverage from the intra-aperture ($B<8$ m) baselines, combined with the high resolution of the inter-aperture ($B>8$ m) baselines allows us to robustly image MWC 297 on scales down to $\sim6$ au.

\section{Observations}\label{sec:obs}

We observed MWC 297 and a reference PSF calibrator on May 31, 2018 using LBTI/LMIRCam \citep[e.g.][]{2012SPIE.8446E..4FL} and the 12-hole non-redundant mask.
We chose the calibrator HD 164259 due to its proximity on the sky to MWC 297, and its similar 2-4 $\mu$m fluxes (Table \ref{tab:targs}). 
HD 164259 is significantly brighter than MWC 297 at the wavefront sensing band (600-900 nm), and slightly fainter than MWC 297 at the phase-tracking band (Ks-band). 
As a result, the calibrator observations had superior AO correction, but poorer co-phasing performance. 
We discuss this further in Appendix \ref{sec:avg}, where we also present data reduction strategies for this situation.

\begin{figure}[ht]
\begin{tabular}{c} 
\includegraphics[width=0.4\textwidth]{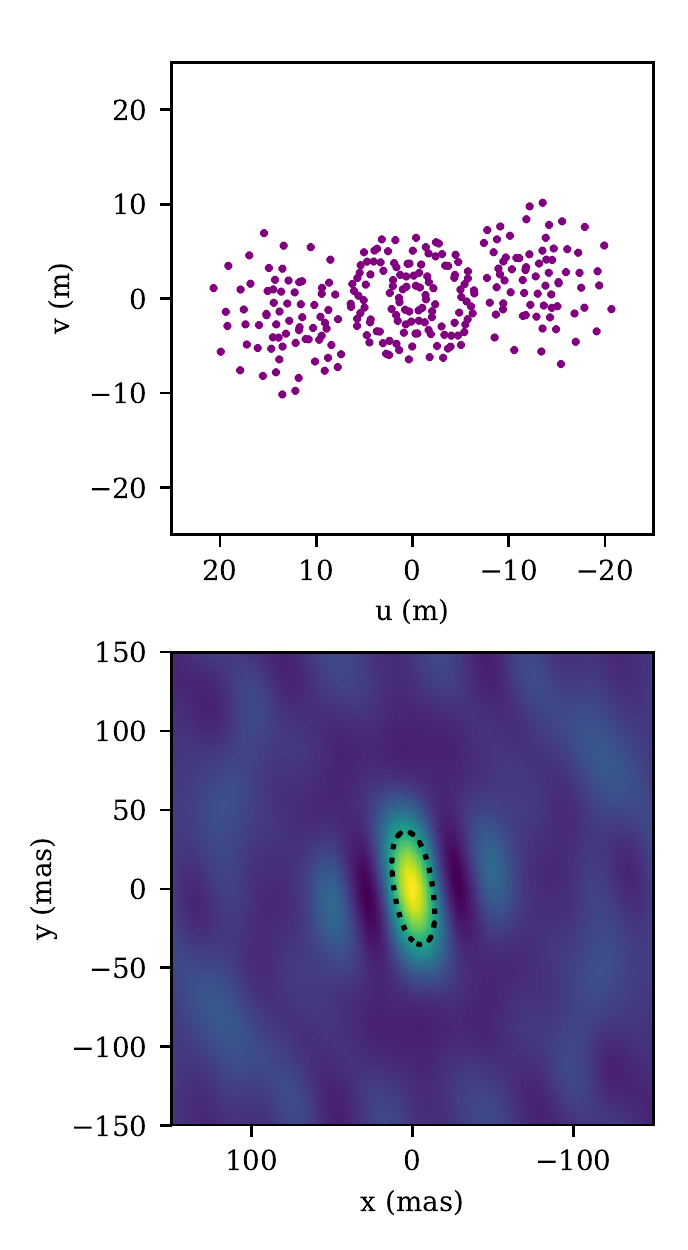}
\end{tabular}
\caption
{ \label{fig:uvcov}
Top: Scattered points show (u,v) coverage for the two MWC 297 pointings (north up, east left). Bottom: Synthesized beam for the (u,v) coverage shown in the top panel (north up, east left). The dotted contour shows 50\% of the peak flux.
}
\end{figure}

We alternated pointings between MWC 297 and HD 164259, using the same integration time ($\sim0.15$ s) for both (Table \ref{tab:tints}).
Each pointing was split into two dithers on the upper and lower halves of the detector.
To fill in the (u,v) plane, we observed in pupil-stabilized mode, with the sky rotating on the detector throughout the night.
Altogether we obtained two pointings of each object, with a total of $\sim19^\circ$ of parallactic angle evolution and $\sim500$s of integration time for MWC 297.
Figure \ref{fig:uvcov} shows the combined (u,v) coverage for the two MWC 297 pointings.

\begin{deluxetable}{lccccc}
\tabletypesize{\footnotesize}
\tablecaption{ Targets \label{tab:targs}}
\tablehead{
\colhead{Name} &  \colhead{RA} & \colhead{DEC} &  \colhead{R} & \colhead{K} & \colhead{L} \\
\colhead{} & \colhead{(hh:mm:ss.ss)} & \colhead{(dd:mm:ss.ss)} & \colhead{(mag)} & \colhead{(mag)} & \colhead{(mag)}
}
\startdata
MWC297 & 18 27 39.53 & -03 49 52.14 & 11.34 & 3.04 & 1.17\\
HD 164259 & 18 00 29.01 & -03 41 24.97 & 4.29 & 3.64 & 3.71 \\
\enddata
\end{deluxetable}

\begin{deluxetable}{lccccc}
\tabletypesize{\footnotesize}
\tablecaption{ Observations: May 31, 2018 \label{tab:tints}}
\tablehead{
\colhead{Pointing} &  \colhead{$n_{frames}$} &  \colhead{$t_{tot}$ (s)} & \colhead{$\overline{seeing}~('')$} & \colhead{UT$_{start}$} & \colhead{UT$_{end}$} 
}
\startdata
HD164259 1 & 1991  & 291 & 0.96 & 07:41:58 & 08:12:32 \\
MWC 297 1 & 1724 & 252 & 0.92 & 08:39:40 & 09:04:11 \\
HD164259 2 & 1997 & 292 & 0.84 & 09:09:48 & 09:27:57\\
MWC 297 2 & 1597 & 233 & 0.90 & 09:33:07 & 10:06:09\\
\enddata
\end{deluxetable}

\section{Data Reduction}\label{sec:red}

\begin{figure}[ht]
\begin{tabular}{c} 
\includegraphics[width=0.4\textwidth]{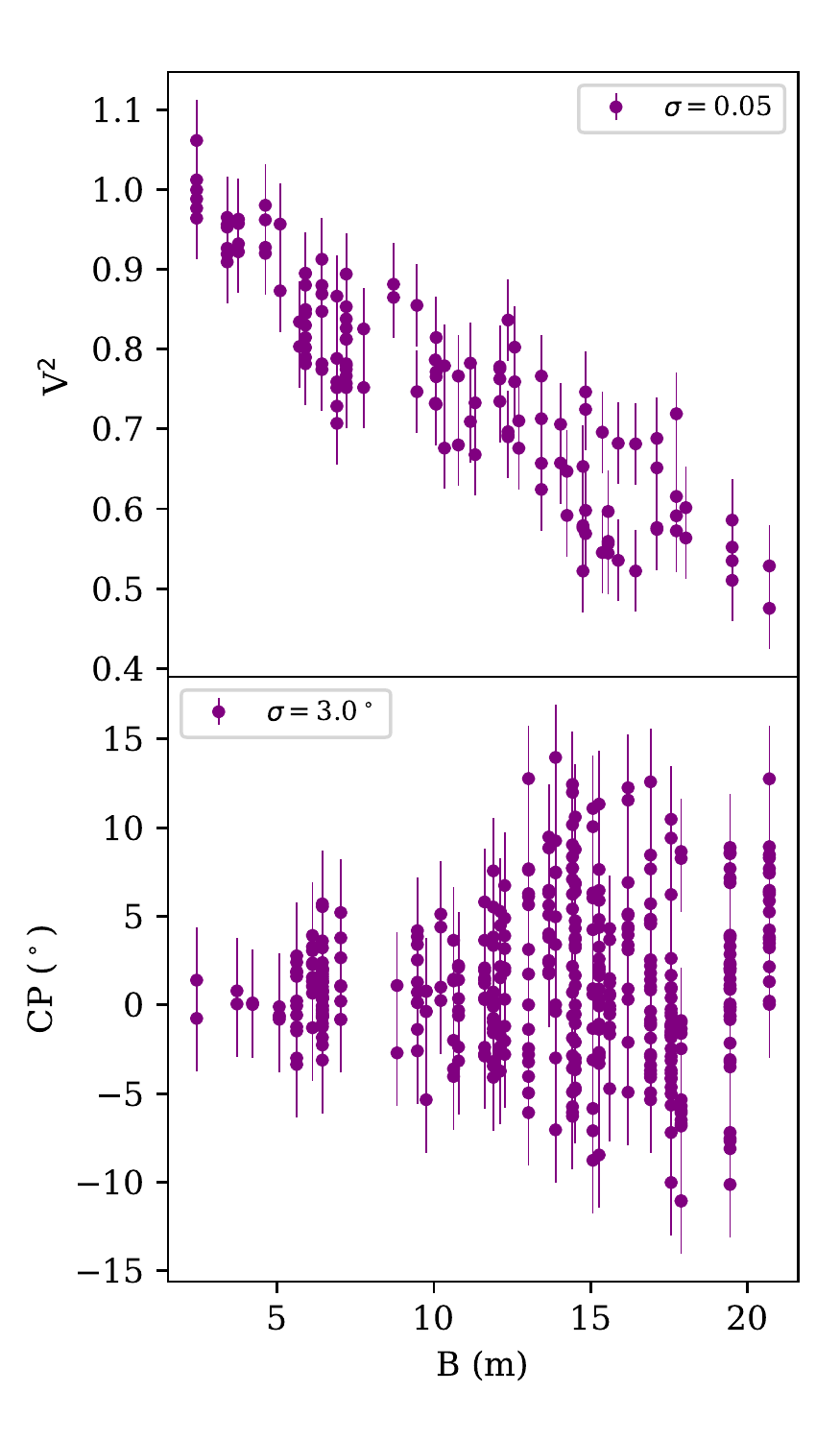}
\end{tabular}
\caption
{ \label{fig:finalobs}
Final calibrated squared visibility versus baseline length (top) and closure phase (bottom) versus maximum triangle baseline length. The raw data were combined using the $p=3$ averaging scheme described in Appendix \ref{sec:avg}, and error bars are assigned following Appendix \ref{sec:errors}.
}
\end{figure}

We reduce the data using an updated version of the pipeline presented in \citet{2017ApJS..233....9S}.
This applies image calibrations, generates Fourier sampling coordinates for the mask, extracts squared visibilities and closure phases, and calibrates the target observables using reference PSF observations. 
Appendix \ref{app:red} provides details on these steps, a description of error bar estimation for the observables, and differences between this reduction and the pipeline presented in \citet{2017ApJS..233....9S}.
Figure \ref{fig:finalobs} shows the final, calibrated visibilities and closure phases.

\section{Analysis}\label{sec:ana}

We reconstruct images using BSMEM \citep{1994IAUS..158...91B,2008SPIE.7013E..3XB}, a gradient-descent algorithm with maximum entropy regularization \citep{1972JOSA...62..511F}. 
For all reconstructions, we use a pixel scale of 1 mas and a field-of-view of 800 mas. 
Changing these parameters does not change the images significantly, unless the field of view is made small compared to the resolution of the shortest baselines.
In this case BSMEM cannot effectively use information from the short baselines.
We run BSMEM in ``classic Bayesian" mode, which chooses the entropy hyperparameter ($\alpha$) automatically by assigning it a prior and evaluating the evidence to choose the most likely value \citep{2008SPIE.7013E..3XB}.
We also test other methods for choosing $\alpha$, which are presented and discussed in Appendix \ref{app:imrecon}.

BSMEM can reconstruct images with a variety of built-in priors or with a user-specified prior image. 
We first reconstruct images with a generic, but physically motivated prior for imaging circumstellar material: a central compact component representing the star, and an extended component to allow for circumstellar structure in the reconstruction.
Previous observations of MWC 297 suggest that it is well modeled by a $\sim$ few mas compact component and a $\gtrsim40$ mas extended component \citep[e.g.][]{2008A&A...485..209A,2007A&A...464...43M,2011A&A...527A.103W}.
We thus begin with a prior image consisting of two circular Gaussian functions - one central, compact component ($FWHM = 2$ mas, fractional flux of $\sim0.8$) and one more extended component ($FWHM = 50$ mas).
We then demonstrate that the prior choice does not significantly change the recovered image by testing several other priors informed by geometric model fits to the data (Appendices \ref{app:modeling} and \ref{app:imrecon}).

\section{Results}\label{sec:results}

\begin{figure}[ht]
\begin{tabular}{c} 
\includegraphics[width=0.48\textwidth]{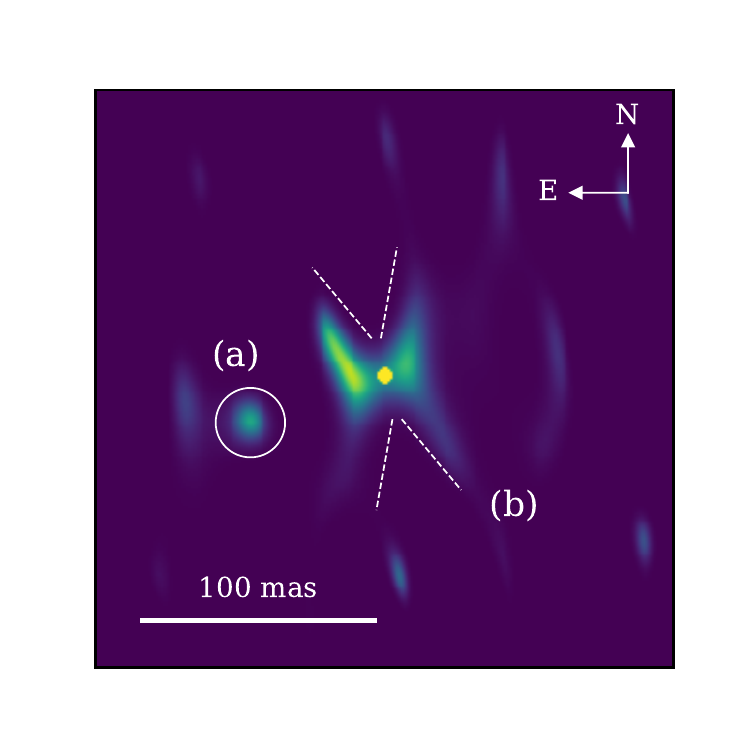}
\end{tabular}
\caption
{ \label{fig:finalim}
Reconstructed image of MWC 297, annotated to highlight: (a) a companion candidate, and (b) a possible outflow cavity (see also dashed lines). The image is shown north up, east left. We discuss these features in Section \ref{sec:disc}. The companion detected in \citet{2020ApJ...890L...8U} lies roughly three image widths to the east. 
}
\end{figure}

\begin{figure}[ht]
\begin{tabular}{c} 
\includegraphics[width=0.4\textwidth]{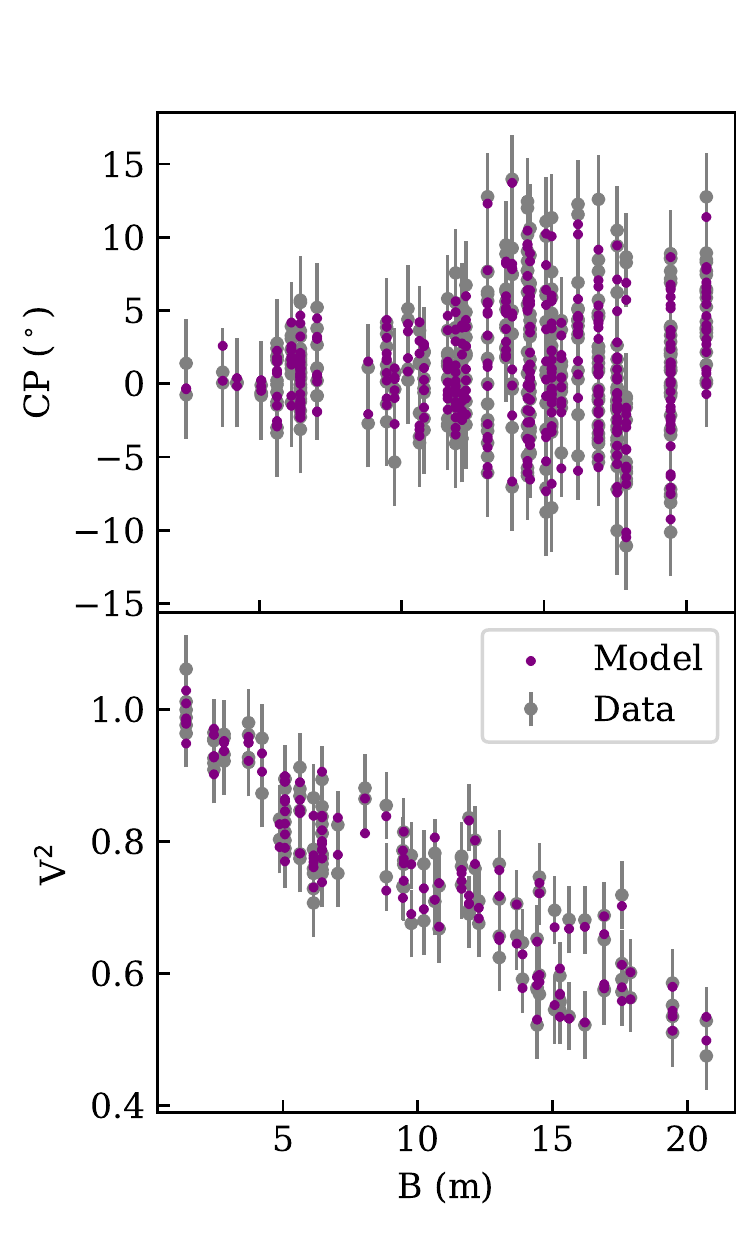}
\end{tabular}
\caption
{ \label{fig:finalimobs}
Purple points show reconstructed closure phases (top) and squared visibilities (bottom) for the image shown in Figure \ref{fig:finalim} as a function of baseline length, plotted over the observations (grey points with error bars).
}
\end{figure}

Figure \ref{fig:finalim} shows the final reconstructed image using the simple two-Gaussian prior, zoomed to the innermost 250 mas, and Figure \ref{fig:finalimobs} shows the reconstructed observables plotted over the data.
MWC 297's morphology is more complex than a simple star plus face-on disk model.
The imaging reveals bright zones along position angles roughly $\pm30^\circ$ east of north, and flux deficits to the north and south (indicated by the dashed lines and (b) label in Figure \ref{fig:finalim}).
The emission to the immediate east of the star is $\sim1.4-2$ times as bright as that to the west.
There is also a companion-like feature (indicated by the circle and (a) label in Figure \ref{fig:finalim}) at $\sim110^\circ$ east of north, with a separation of $\sim60$ mas and contrast of $\sim2\%$ (4.25 mag) relative to the central compact component. 
Appendices \ref{sec:modelres} and \ref{app:simrtdisk} present modeling that assesses the false positive probability of this signal, which we calculate to be $<0.08\%$.

\begin{figure}[ht]
\begin{tabular}{c} 
\includegraphics[width=0.48\textwidth]{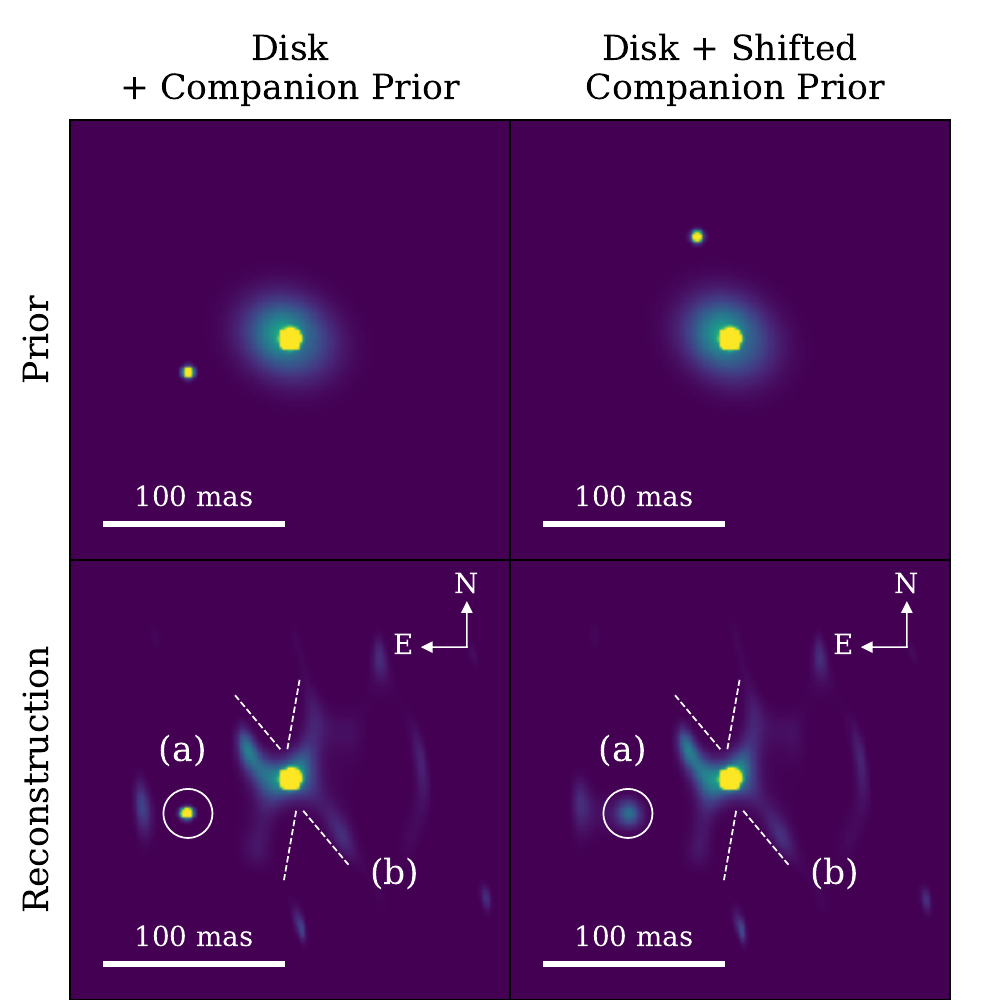}
\end{tabular}
\caption
{ \label{fig:compprior}
BSMEM reconstructions (bottom) using two different priors (top) consisting of a compact central component, an extended disk, and a compact companion. The left column shows the results for a prior where the companion is at the position of the companion signal in Figure \ref{fig:finalim}. The right shows the results where the companion prior position angle is offset by 90$^\circ$. The reconstructed images color scale and annotations are the same as those in Figure \ref{fig:finalim}.
}
\end{figure}

We carried out a variety of tests to demonstrate that the companion signal is independent of prior image choice. 
Figure \ref{fig:compprior} shows one of these, where we reconstruct images using priors made up of a central, compact component, an extended disk, and a compact companion signal. 
In one prior, the companion location is the same as the companion signal in Figure \ref{fig:finalim}, and in the other the position angle is offset.
While only one example is shown in Figure \ref{fig:compprior}, we tested a variety of prior companion position angle offsets.
When the prior companion is at the same position as the signal in Figure \ref{fig:finalim}, the reconstructed companion becomes more compact but contains the same fractional flux.
For the offset prior companions, BSMEM does not introduce fake signals with significant fractional flux compared to either the observed companion or the noise levels in the data.
The images and companion signal are robust to arbitrary prior choices such as these and others that are presented in Appendix \ref{app:prior}.

Appendix \ref{app:imrecon} presents the detailed results of additional image fidelity tests related to error bars and regularization, and we describe them briefly here.
Reconstructions with scaled error bars and alternative entropy hyperparameters ($\alpha$) show that small errors and/or small $\alpha$ lead BSMEM to introduce high frequency peaks throughout the field of view, improving its ability to perfectly match the observations (including noise).
Conversely, large error bars and/or large $\alpha$ result in a greater mismatch between BSMEM's observables and the data.
Varying $\alpha$ and the error bars within reason does not qualitatively change the structure in the innermost 250 mas of the reconstruction (Appendices \ref{app:lcurve} and \ref{app:errs}).

Over-regularized images (which closely resemble face-on disks) do not reproduce the observations even qualitatively.
Furthermore, simulated reconstructed images of $\delta$ functions plus various disk models cannot reproduce the structure seen in Figure \ref{fig:finalim} (Appendix \ref{app:simrecon}).
Gapped radiative transfer disk models (which are inconsistent with previous datasets but which we explore in Appendix \ref{app:simrtdisk}), also cannot reliably reproduce the reconstructed image.
The images thus demonstrate that MWC 297's circumstellar emission is inconsistent with a simple, relatively face-on disk. 
Geometric modeling of the observations support this as well, with high reduced $\chi^2$ values for all disk models, but better reduced $\chi^2$ and Bayesian evidence values for increasingly complex geometric models (e.g.~skewed disks, disks plus companions; Appendix \ref{app:modeling}).

We note that some of the model squared visibilities in Figure \ref{fig:finalimobs} reach values slightly greater than 1. 
This is because BSMEM allows for an error on the zero spacing flux, which we left at its default value of 0.1 for this reconstruction. 
As a result, during BSMEM's automatic regularization, the data have enough weight compared to the regularizer that BSMEM introduces noise to try to better match the visibilities, which reach values greater than 1 due to imperfect calibration. 
This effect diminishes with increasing regularizer values (see Appendix \ref{app:imrecon}, Figures \ref{fig:Lpimages} and \ref{fig:hyper}) and decreasing values of the zero spacing flux error.
It also does not significantly change the morphology in the central $\sim250$ mas of the reconstruction, which can be seen by comparing the reconstruction in Figure \ref{fig:finalim} to those in Figures \ref{fig:Lpimages} and \ref{fig:hyper}.

\section{Discussion}\label{sec:disc}

\subsection{MWC 297's Circumstellar Disk}\label{subsec:disc:disk}
Here we use the reconstructed images to place constraints on MWC 297's circumstellar structure. 
The imaging reveals an unresolved central component with a large fractional flux; 0.74-0.78 on few mas scales.
We use MWC 297's spectral energy distribution to estimate the fractional flux that can be associated with disk material.
Assuming a stellar effective temperature and radius of $\sim23,700$ K, and $\sim9.17~\mathrm{R_\odot}$, respectively \citep{2020ApJ...890L...8U}, the stellar contribution to MWC 297's total 3.7 $\mu$m flux is $\sim5\%$.
This suggests that $\sim69-72\%$ of the total 3.7 $\mu$m flux resides in a compact region that is unresolved by the LBTI observations. 
This is consistent with recent VLTI modeling and imaging, which showed that the $1.65-2.2~\mu$m stellar fractional flux was $\sim10-15\%$, and that the remainder was concentrated in an area with a characteristic size of a few mas \citep[e.g.][]{2007A&A...464...43M,2020A&A...636A.116K}.

We explore whether the remaining extended emission can be explained by the relatively face-on disk scenario suggested by previous VLTI observations \citep[e.g.][]{2020A&A...636A.116K}
For models of this type, the LBTI data prefer inner radii on the scale of several au; we thus test gapped disk models that can roughly match our observations without underestimating the VLTI squared visibilities or closure phases presented in \citet{2020A&A...636A.116K}.
These tests show that gapped disk models significantly under-fit the LBTI closure phases, do not reproduce the reconstructed image in the absence of noise, and do not reliably reproduce the reconstructed image with appropriate levels of added noise (see Appendix \ref{app:simrtdisk}).

The remaining flux around MWC 297 is thus more complex than a simple, face-on disk with or without gaps.
The central $\sim3-7$ au region of the reconstructed image is relatively symmetric, but further out ($\sim10-30$ au) the flux is distributed in a butterfly pattern similar to near-infrared HST images of protostars with outflows \citep[e.g.][]{1999AJ....117.1490P,2001ApJ...556..958C,2008ApJ...674L.101W,1996ApJ...473..437B}.
Herbig Ae/Be stars are thought to drive well-collimated outflows (opening angle $\lesssim50^\circ$) at ages of less than a few $\times~10^4$ yr, and poorly-collimated outflows at older ages \citep[e.g][]{2005ASSL..324..105B}.
Given MWC 297's young age ($\sim2.8\times10^4$ yr) one may expect relatively collimated outflows, which could cause this disk morphology.
The nearly-vertical, conical dark zone in the reconstructed image could be caused by a well-collimated bipolar outflow, which would agree with the north-south elongation seen in a previously-published 5 GHz map \citep{1997MNRAS.286..538D}.

The inner extent of the possible outflow cavity ($\sim7$ au) is consistent with the inner reaches of outflows observed around Herbig stars \citep[e.g.][]{2000ApJ...542L.115D}.
The spatial scale of the entire butterfly pattern ($\sim30$ au north-south, $\sim20$ au east-west) is smaller than that observed in the infrared for protostars with outflow cavities ($\gtrsim100$ au).
It is possible that the LBTI observations trace only the brightest regions of the disk and either would not be sensitive to, or would over-resolve fainter, more extended emission.
It is also possible that MWC 297's higher mass, and thus higher photoionizing flux, leads to a more compact disk compared to the T Tauri stars where these nebulae have been observed.
Lastly, a companion with a separation of tens of au (comparable to the companion candidate separation in the reconstructed image) could also truncate the disk, leading to the more compact emission seen here.

The observed brightness distribution is most similar to disk plus outflow models viewed at moderate inclination \citep[$\gtrsim50-60^\circ$; e.g][]{2006ApJ...649..900S}.
This explanation requires that the disk inclination be higher than the $\sim15-38^\circ$ estimated from previous long-baseline observations \citep[e.g.][]{2001ApJ...546..358M,2004ApJ...613.1049E,2006ApJ...647..444M,2007A&A...464...43M,2008A&A...485..209A,2011A&A...527A.103W,2017A&A...607A..17H,2017A&A...599A..85L,2020A&A...636A.116K}.
For all of these except \citet{2020A&A...636A.116K}, the inclination estimates were based on the aspect ratios of geometric fits to visibilities.
\citet{2020A&A...636A.116K} based their inclination estimate ($\sim38^\circ$) on a disk fit that reproduced the location and degree of the asymmetry in the reconstructed image.
However, more complex structures than simple face-on disks may have relatively symmetric brightness distributions at such small scales, causing geometric and disk model fits to prefer face-on geometries.
Indeed, the inner regions of simulated disks with outflow cavities have been shown to appear relatively symmetric for moderate inclinations \citep[$\lesssim65^\circ$; e.g.][]{{2006ApJ...649..900S}}.

We note that the models shown in \citet{2006ApJ...649..900S} are not entirely analogous to MWC 297, since they cannot account for the compact, high fractional flux suggested by the LBT and long-baseline observations. 
Recent VLTI observations do not show a decrease in brightness toward small radii, suggesting they do not resolve the inner radius of the circumstellar emission \citep{2020A&A...636A.116K}.
Furthermore, fits to previous VLTI datasets and MWC 297's spectral energy distribution prefer disk models with continuum emission inside the sublimation radius \citep[e.g.][]{2008A&A...485..209A,2007A&A...464...43M,2011A&A...527A.103W}. 
This suggests either the presence of free-free emission or refractory grains close to the star, neither of which are taken into account in \citet{2006ApJ...649..900S}.
While fully exploring such a complex system's parameter space is beyond the scope of this paper, either of these scenarios should still produce enough near-infrared radiation to cause extended emission from scattering by an outflow cavity.

The disk plus outflow scenario we propose here implies a higher inclination than the $38^\circ$ suggested by previous long baseline studies.
If this were the case, one might expect to see close-in asymmetric emission on one side of the star from an inner disk rim located at $\sim7$ au (the dust sublimation radius).
However, the presence of free-free emission or close-in refractory grains could result in a non detection of an asymmetric disk rim. 
In both cases the bright central component could increase the contrast of a close-in disk, reducing its closure phase signal. 
A refractory inner disk could also shadow a rim of non-refractory grains, increasing its contrast. 
Lastly, the higher inclination implied by the disk plus outflow scenario would be consistent with the observed $v\sin{i}$ of 350 km/s, which requires an $i\gtrsim50^\circ$ for the rotational velocity to be less than breakup ($\sim480$ km/s).
Future comprehensive modeling efforts could nail down the validity of this scenario and possibly measure the true inclination of MWC 297.

\vspace{-10pt}\subsection{A Companion Candidate Around MWC 297}\label{subsec:disc:comp}
We detect a companion candidate around MWC 297 at a separation of $\sim60$ mas and a contrast of $\sim2\%=4.25$ mag relative to the unresolved flux.
The observed contrast agrees with previous near-infrared visual binary searches that observed a trend of sharply increasing companion contrasts greater than $\Delta K\sim2$ mag \citep[e.g.][]{1997A&A...318..472L}.
The projected spatial separation ($\sim22$ au at a distance of 375 pc) lies in a region of the parameter space where aperture masking and long-baseline interferometry studies have collectively set a lower limit of $\sim11\%$ on the companion fraction, with detected companions having spatial separations of $\sim1.4-30$ au \citep[e.g.][]{2015A&A...574A..41A,2005A&A...431..307S,2012ApJ...746L...2K,2011A&A...529L...1B,2015Ap&SS.355..291D}.
Here we discuss possible physical explanations for this companion signal.

The companion candidate fractional flux at 3.7$\mu$m is $1.5\pm0.2\%$, corresponding to an intrinsic flux of $\sim1.12\pm0.17$ Jy.
Assuming the same age as MWC 297 \citep[$\sim2.8\times10^4$ yr;][]{2018A&A...620A.128V}, and that the emission comes from a bare photosphere, we use the solar-metallicity PARSEC \citep{2012MNRAS.427..127B} pre-main-sequence evolutionary tracks to fit for a stellar mass, radius, and effective temperature.
Models with $\mathrm{M_*} = 2.1\pm^{0.2}_{0.3}~\mathrm{M_\odot}$, $\mathrm{R_*} = 13.1\pm^{0.7}_{1.4}~\mathrm{R_\odot}$, and $\mathrm{T_{eff}} = 4542\pm^{150}_{40}~\mathrm{K}$ are consistent with the 3.7$\mu$m flux.
A stellar companion such as this one would not be noticeable in previously-published H band to K band long-baseline interferometry visibilities.
However, it may be noticeable in the H band closure phases published in \citet{2020A&A...636A.116K}, since the expected H band flux is $0.7-0.9$ Jy after reddening.

To check this, we generate a $\delta$ + skewed Gaussian model image with the same stellar fractional flux, FWHM, and asymmetry as the reconstructed image in \citet{2020A&A...636A.116K}.
We sample it with the same ($u,v$) coverage as \citet{2020A&A...636A.116K}, and compare the results for this image to one where we add a companion candidate with a flux of 0.8 Jy (corresponding to a contrast of $\sim15\%$) and a 60 mas separation.
The squared visibilities for these two cases are indistinguishable, but the closure phases for the model with the companion are on average $8^\circ$ larger than the $\delta$ + skewed Gaussian model, much larger than the typical error bars.

The inconsistency between a bare stellar photosphere and the previously-published VLTI closure phases suggests that the companion must be very red.
To explore this, we increase the contrast of the companion in the model images described above, until the difference between the $\delta$ + skewed Gaussian model and the $\delta$ + skewed Gaussian + companion model is smaller than the typical VLTI closure phase error bar.
The change associated with a contrast of $\sim1\%$ with respect to the central star is below these limits, corresponding to an H-band flux of $\sim0.05$ Jy  and an H-L color $> 4.8$ mag (compared to $\sim0.7$ mag for MWC 297).
At an age of $\sim2.8\times10^4$ yr, companions with masses less than $\sim0.1-0.5~\mathrm{M_\odot}$ can satisfy this H-band upper limit, but cannot reproduce the color. 
However, companions surrounded by warm ($\sim700$ K) dust can satisfy both the H and K band flux constraints if their extent is $\gtrsim3.5$ au.
A dust-shrouded companion with a mass $\gtrsim0.2~\mathrm{M_\odot}$ could match this scenario, since its Hill radius would be $\gtrsim3.5$ au for an orbital radius of $\sim22$ au.

Single epoch observations cannot distinguish between an orbiting, dusty companion and circumstellar material. 
MWC 297's mass ($\sim17~\mathrm{M_\odot}$) and the companion separation  ($22$ au) imply an orbital period of at least $\sim25$ years and evolution of up to $\sim14^\circ$ yr$^{-1}$ in position angle.
Given the companion position angle uncertainty of a few degrees, followup observations in the coming years should be capable of observing or ruling out the expected orbital motion at high significance.
An alternative explanation with no expected orbital motion is that the companion signal originates from a Herbig Haro object \citep{1951ApJ...113..697H,1952ApJ...115..572H} caused by an equatorial flow (assuming the north-south cavity in the imaging is caused by a bipolar flow). 
Modeling of radiation-driven winds has shown that both bipolar and equatorial outflows may exist around accreting early Herbig Be stars \citep[e.g.][]{1998MNRAS.296L...6D}.

\subsection{3.7$\mu$m Constraints on the Wide-Separation SPHERE Companion}\label{subsec:disc:sphere}
VLT/SPHERE recently detected a companion around MWC 297 at a separation of $\sim650$ mas, corresponding to $\sim245$ au \citep{2020ApJ...890L...8U}.  
We do not detect significant emission at the SPHERE companion location in the reconstructed images.
Given the noise levels in the observations, the simulations presented in Appendix \ref{app:modeling} suggest that the L$'$ companion contrast is greater than $\sim5$ magnitudes, corresponding to a flux less than $\sim750$ mJy.
This is consistent with the best-fit BT-SETTL model presented in \citet{2020ApJ...890L...8U}, which predicts a $\sim6$ mJy 3.7 $\mu$m flux ($\Delta L' \sim 10$ mag).

\section{Conclusions}\label{sec:conc}
We presented new, spatially resolved observations of the Herbig Be star MWC 297 from the co-phased LBTI.
The imaging allows us to characterize MWC 297's disk geometry in detail.
It reveals complex disk structure on scales of $\sim10-30$ au that closely resembles butterfly patterns associated with collimated outflows around young stars.
Protostellar outflow models with moderate inclinations ($\sim50-65^\circ$) provide a good match to the data.
This scenario is consistent with MWC 297's young age and spectral type, since early Be stars are expected to drive collimated outflows at ages $\lesssim$ a few $\times~10^4$ yr.

The images resolve inconsistencies between low inclination estimates and high stellar $v\sin{i}$ measurements in previous studies.
Previous inclination constraints ($\sim15-38^\circ$) were all based on simple disk models to primarily long-baseline visibility data. 
Indeed, the axes ratios of simple geometric disk fits to the LBTI data also imply a low inclination (Appendix \ref{app:modeling}).
However, the imaging may be better explained by a moderately inclined disk plus outflow, a scenario that has been demonstrated to look relatively symmetric through radiative transfer modeling. 
This larger inclination ($\sim50-65^\circ$) would be consistent with the large $v\sin{i}$ ($350~\mathrm{km~s^{-1}}$), which requires $i\gtrsim50^\circ$ for the stellar rotation to be below breakup.
While fully modeling this complex system is beyond the scope of this paper, future modeling efforts could confirm or refute this scenario and possibly measure the true inclination of MWC 297.

The images also show a companion-like feature at a separation of 60 mas, corresponding to 22 au at the distance of MWC 297.
If this feature is indeed an orbiting companion, its infrared flux constraints cannot be explained by a bare stellar photosphere. 
They are better matched by warm $\sim700$ K dust on the scales of a few au, corresponding to the size of the Hill sphere for companions with $\mathrm{M_*} \gtrsim 0.2~\mathrm{M_\odot}$. 
Multi-epoch observations will distinguish between a dusty companion scenario and alternatives, such as heating of disk material by an equatorial outflow. 

The images of MWC 297 have $\sim17$ mas resolution at 3.7 $\mu$m.
Unlike previous LBTI NRM datasets taken in ``lucky'' imaging mode \citep{2017ApJ...844...22S}, here we demonstrate that the structure in the reconstructed images is independent of prior choices, regularization hyperparameters, and issues in error bar estimation.
The imaging data provide a high-fidelity, ELT-resolution view of several au scales around a distant young star. 
While studies applying geometric models provide some constraints on MWC 297's circumstellar environment, we demonstrate that robust imaging is required to characterize its morphology in detail and resolve modeling inconsistencies.
Future observations with the 23-m LBTI and with upcoming facilities such as GMT and TMT will place new, similarly detailed constraints on disk structures and companions around young stellar objects like MWC 297.

\acknowledgements
S.S. acknowledges support from NSF award number 1701489.
J.A.E. acknowledges support from NSF award number 1745406 and NASA award NNX16AJ74G.
J.M.S. is supported by NASA through Hubble Fellowship grant HST-HF2-51398.001-A awarded by the Space Telescope Science Institute, which is operated by the Association of Universities for Research in Astronomy, Inc., for NASA, under contract NAS5-26555.

The LBT is an international collaboration among institutions in the United States,Italy, and Germany. LBT Corporation partners are The University of Arizona on behalf of the Arizona university system; Istituto Nazionale di Astrofisica, Italy; LBT Beteiligungsgesellschaft, Germany, representing the Max-Planck Society, the Astrophysical Institute  Potsdam,  and  Heidelberg  University;  The  Ohio State University, and The Research Corporation, on behalf of The University of Notre Dame, University of Minnesota, and University of Virginia.

\bibliography{references2}

\begin{thebibliography}{}
\expandafter\ifx\csname natexlab\endcsname\relax\def\natexlab#1{#1}\fi
\providecommand{\url}[1]{\href{#1}{#1}}
\providecommand{\dodoi}[1]{doi:~\href{http://doi.org/#1}{\nolinkurl{#1}}}
\providecommand{\doeprint}[1]{\href{http://ascl.net/#1}{\nolinkurl{http://ascl.net/#1}}}
\providecommand{\doarXiv}[1]{\href{https://arxiv.org/abs/#1}{\nolinkurl{https://arxiv.org/abs/#1}}}

\bibitem[{{Acke} {et~al.}(2008){Acke}, {Verhoelst}, {van den Ancker}, {Deroo},
  {Waelkens}, {Chesneau}, {Tatulli}, {Benisty}, {Puga}, {Waters}, {Verhoeff},
  \& {de Koter}}]{2008A&A...485..209A}
{Acke}, B., {Verhoelst}, T., {van den Ancker}, M.~E., {et~al.} 2008, \aap, 485,
  209, \dodoi{10.1051/0004-6361:200809654}

\bibitem[{{Alonso-Albi} {et~al.}(2009){Alonso-Albi}, {Fuente}, {Bachiller},
  {Neri}, {Planesas}, {Testi}, {Bern{\'e}}, \& {Joblin}}]{2009A&A...497..117A}
{Alonso-Albi}, T., {Fuente}, A., {Bachiller}, R., {et~al.} 2009, \aap, 497,
  117, \dodoi{10.1051/0004-6361/200810401}

\bibitem[{{Anthonioz} {et~al.}(2015){Anthonioz}, {M{\'e}nard}, {Pinte}, {Le
  Bouquin}, {Benisty}, {Thi}, {Absil}, {Duch{\^e}ne}, {Augereau}, {Berger},
  {Casassus}, {Duvert}, {Lazareff}, {Malbet}, {Millan-Gabet}, {Schreiber},
  {Traub}, \& {Zins}}]{2015A&A...574A..41A}
{Anthonioz}, F., {M{\'e}nard}, F., {Pinte}, C., {et~al.} 2015, \aap, 574, A41,
  \dodoi{10.1051/0004-6361/201424520}

\bibitem[{{Bailey} {et~al.}(2014){Bailey}, {Hinz}, {Puglisi}, {Esposito},
  {Vaitheeswaran}, {Skemer}, {Defr{\`e}re}, {Vaz}, \&
  {Leisenring}}]{2014SPIE.9148E..03B}
{Bailey}, V.~P., {Hinz}, P.~M., {Puglisi}, A.~T., {et~al.} 2014, in \procspie,
  Vol. 9148, Adaptive Optics Systems IV, 914803, \dodoi{10.1117/12.2057138}

\bibitem[{{Baines} {et~al.}(2006){Baines}, {Oudmaijer}, {Porter}, \&
  {Pozzo}}]{2006MNRAS.367..737B}
{Baines}, D., {Oudmaijer}, R.~D., {Porter}, J.~M., \& {Pozzo}, M. 2006, \mnras,
  367, 737, \dodoi{10.1111/j.1365-2966.2006.10006.x}

\bibitem[{{Bans} \& {K{\"o}nigl}(2012)}]{2012ApJ...758..100B}
{Bans}, A., \& {K{\"o}nigl}, A. 2012, \apj, 758, 100,
  \dodoi{10.1088/0004-637X/758/2/100}

\bibitem[{{Baron} \& {Young}(2008)}]{2008SPIE.7013E..3XB}
{Baron}, F., \& {Young}, J.~S. 2008, in Society of Photo-Optical
  Instrumentation Engineers (SPIE) Conference Series, Vol. 7013, Society of
  Photo-Optical Instrumentation Engineers (SPIE) Conference Series, 3,
  \dodoi{10.1117/12.789115}

\bibitem[{{Berger} {et~al.}(2011){Berger}, {Monnier}, {Millan-Gabet}, {Renard},
  {Pedretti}, {Traub}, {Bechet}, {Benisty}, {Carleton}, {Haguenauer}, {Kern},
  {Labeye}, {Longa}, {Lacasse}, {Malbet}, {Perraut}, {Ragland}, {Schloerb},
  {Schuller}, \& {Thi{\'e}baut}}]{2011A&A...529L...1B}
{Berger}, J.~P., {Monnier}, J.~D., {Millan-Gabet}, R., {et~al.} 2011, \aap,
  529, L1, \dodoi{10.1051/0004-6361/201016219}

\bibitem[{{Bergner} {et~al.}(1988){Bergner}, {Kozlov}, {Krivtsov},
  {Miroshnichenko}, {Yudin}, {Yutanov}, {Dzhakusheva}, {Kuratov}, \&
  {Mukanov}}]{1988Ap.....28..313B}
{Bergner}, Y.~K., {Kozlov}, V.~P., {Krivtsov}, A.~A., {et~al.} 1988,
  Astrophysics, 28, 313, \dodoi{10.1007/BF01112966}

\bibitem[{{Beuther} \& {Shepherd}(2005)}]{2005ASSL..324..105B}
{Beuther}, H., \& {Shepherd}, D. 2005, Astrophysics and Space Science Library,
  Vol. 324, {Precursors of UchII Regions and the Evolution of Massive
  Outflows}, ed. M.~S.~N. {Kumar}, M.~{Tafalla}, \& P.~{Caselli}, 105,
  \dodoi{10.1007/0-387-26357-8_8}

\bibitem[{{Bouvier} \& {Corporon}(2001)}]{2001IAUS..200..155B}
{Bouvier}, J., \& {Corporon}, P. 2001, in IAU Symposium, Vol. 200, The
  Formation of Binary Stars, ed. H.~{Zinnecker} \& R.~{Mathieu}, 155

\bibitem[{{Bressan} {et~al.}(2012){Bressan}, {Marigo}, {Girardi}, {Salasnich},
  {Dal Cero}, {Rubele}, \& {Nanni}}]{2012MNRAS.427..127B}
{Bressan}, A., {Marigo}, P., {Girardi}, L., {et~al.} 2012, \mnras, 427, 127,
  \dodoi{10.1111/j.1365-2966.2012.21948.x}

\bibitem[{{Burrows} {et~al.}(1996){Burrows}, {Stapelfeldt}, {Watson}, {Krist},
  {Ballester}, {Clarke}, {Crisp}, {Gallagher}, {Griffiths}, {Hester},
  {Hoessel}, {Holtzman}, {Mould}, {Scowen}, {Trauger}, \&
  {Westphal}}]{1996ApJ...473..437B}
{Burrows}, C.~J., {Stapelfeldt}, K.~R., {Watson}, A.~M., {et~al.} 1996, \apj,
  473, 437, \dodoi{10.1086/178156}

\bibitem[{{Buscher}(1994)}]{1994IAUS..158...91B}
{Buscher}, D.~F. 1994, in IAU Symposium, Vol. 158, Very High Angular Resolution
  Imaging, ed. J.~G. {Robertson} \& W.~J. {Tango}, 91--93

\bibitem[{{Canto} {et~al.}(1984){Canto}, {Rodriguez}, {Calvet}, \&
  {Levreault}}]{1984ApJ...282..631C}
{Canto}, J., {Rodriguez}, L.~F., {Calvet}, N., \& {Levreault}, R.~M. 1984,
  \apj, 282, 631, \dodoi{10.1086/162242}

\bibitem[{{Corporon} \& {Lagrange}(1999)}]{1999A&AS..136..429C}
{Corporon}, P., \& {Lagrange}, A.~M. 1999, \aaps, 136, 429,
  \dodoi{10.1051/aas:1999225}

\bibitem[{{Cotera} {et~al.}(2001){Cotera}, {Whitney}, {Young}, {Wolff}, {Wood},
  {Povich}, {Schneider}, {Rieke}, \& {Thompson}}]{2001ApJ...556..958C}
{Cotera}, A.~S., {Whitney}, B.~A., {Young}, E., {et~al.} 2001, \apj, 556, 958,
  \dodoi{10.1086/321627}

\bibitem[{{Delfosse} \& {Bonneau}(2004)}]{2004sf2a.conf..181D}
{Delfosse}, X., \& {Bonneau}, D. 2004, in SF2A-2004: Semaine de l'Astrophysique
  Francaise, ed. F.~{Combes}, D.~{Barret}, T.~{Contini}, F.~{Meynadier}, \&
  L.~{Pagani}, 181

\bibitem[{{Devine} {et~al.}(2000){Devine}, {Grady}, {Kimble}, {Woodgate},
  {Bruhweiler}, {Boggess}, {Linsky}, \& {Clampin}}]{2000ApJ...542L.115D}
{Devine}, D., {Grady}, C.~A., {Kimble}, R.~A., {et~al.} 2000, \apjl, 542, L115,
  \dodoi{10.1086/312939}

\bibitem[{{Drew} {et~al.}(1997){Drew}, {Busfield}, {Hoare}, {Murdoch}, {Nixon},
  \& {Oudmaijer}}]{1997MNRAS.286..538D}
{Drew}, J.~E., {Busfield}, G., {Hoare}, M.~G., {et~al.} 1997, \mnras, 286, 538,
  \dodoi{10.1093/mnras/286.3.538}

\bibitem[{{Drew} {et~al.}(1998){Drew}, {Proga}, \&
  {Stone}}]{1998MNRAS.296L...6D}
{Drew}, J.~E., {Proga}, D., \& {Stone}, J.~M. 1998, \mnras, 296, L6,
  \dodoi{10.1046/j.1365-8711.1998.01438.x}

\bibitem[{{Duch{\^e}ne}(2015)}]{2015Ap&SS.355..291D}
{Duch{\^e}ne}, G. 2015, \apss, 355, 291, \dodoi{10.1007/s10509-014-2173-7}

\bibitem[{{Dullemond}(2012)}]{2012ascl.soft02015D}
{Dullemond}, C.~P. 2012, {RADMC-3D: A multi-purpose radiative transfer tool},
  Astrophysics Source Code Library.
\newblock \doeprint{1202.015}

\bibitem[{{Eisner} {et~al.}(2004){Eisner}, {Lane}, {Hillenbrand}, {Akeson}, \&
  {Sargent}}]{2004ApJ...613.1049E}
{Eisner}, J.~A., {Lane}, B.~F., {Hillenbrand}, L.~A., {Akeson}, R.~L., \&
  {Sargent}, A.~I. 2004, \apj, 613, 1049, \dodoi{10.1086/423314}

\bibitem[{{Eisner} {et~al.}(2015){Eisner}, {Rieke}, {Rieke}, {Flaherty},
  {Stone}, {Arnold}, {Cortes}, {Cox}, {Hawkins}, {Cole}, {Zajac}, \&
  {Rudolph}}]{2015MNRAS.447..202E}
{Eisner}, J.~A., {Rieke}, G.~H., {Rieke}, M.~J., {et~al.} 2015, \mnras, 447,
  202, \dodoi{10.1093/mnras/stu2441}

\bibitem[{{Foreman-Mackey} {et~al.}(2013){Foreman-Mackey}, {Hogg}, {Lang}, \&
  {Goodman}}]{2013PASP..125..306F}
{Foreman-Mackey}, D., {Hogg}, D.~W., {Lang}, D., \& {Goodman}, J. 2013, \pasp,
  125, 306, \dodoi{10.1086/670067}

\bibitem[{{Frieden}(1972)}]{1972JOSA...62..511F}
{Frieden}, B.~R. 1972, Journal of the Optical Society of America (1917-1983),
  62, 511

\bibitem[{{Goggans} \& {Chi}(2004)}]{2004AIPC..707...59G}
{Goggans}, P.~M., \& {Chi}, Y. 2004, in American Institute of Physics
  Conference Series, Vol. 707, Bayesian Inference and Maximum Entropy Methods
  in Science and Engineering, ed. G.~J. {Erickson} \& Y.~{Zhai}, 59--66,
  \dodoi{10.1063/1.1751356}

\bibitem[{{Guyon} {et~al.}(2014){Guyon}, {Hinz}, {Cady}, {Belikov}, \&
  {Martinache}}]{2014ApJ...780..171G}
{Guyon}, O., {Hinz}, P.~M., {Cady}, E., {Belikov}, R., \& {Martinache}, F.
  2014, \apj, 780, 171, \dodoi{10.1088/0004-637X/780/2/171}

\bibitem[{Hansen(1992)}]{Hansen1992}
Hansen, P.~C. 1992, SIAM Review, 34, 561, \dodoi{10.1137/1034115}

\bibitem[{{Haro}(1952)}]{1952ApJ...115..572H}
{Haro}, G. 1952, \apj, 115, 572, \dodoi{10.1086/145576}

\bibitem[{{Herbig}(1951)}]{1951ApJ...113..697H}
{Herbig}, G.~H. 1951, \apj, 113, 697, \dodoi{10.1086/145440}

\bibitem[{{Herbig}(1960)}]{1960ApJS....4..337H}
---. 1960, \apjs, 4, 337, \dodoi{10.1086/190050}

\bibitem[{{Hillenbrand} {et~al.}(1992){Hillenbrand}, {Strom}, {Vrba}, \&
  {Keene}}]{1992ApJ...397..613H}
{Hillenbrand}, L.~A., {Strom}, S.~E., {Vrba}, F.~J., \& {Keene}, J. 1992, \apj,
  397, 613, \dodoi{10.1086/171819}

\bibitem[{{Hinz} {et~al.}(2008){Hinz}, {Bippert-Plymate}, {Breuninger},
  {Connors}, {Duffy}, {Esposito}, {Hoffmann}, {Kim}, {Kraus}, {McMahon},
  {Montoya}, {Nash}, {Durney}, {Solheid}, {Tozzi}, \&
  {Vaitheeswaran}}]{2008SPIE.7013E..28H}
{Hinz}, P.~M., {Bippert-Plymate}, T., {Breuninger}, A., {et~al.} 2008, in
  Society of Photo-Optical Instrumentation Engineers (SPIE) Conference Series,
  Vol. 7013, Society of Photo-Optical Instrumentation Engineers (SPIE)
  Conference Series, 28--36, \dodoi{10.1117/12.790211}

\bibitem[{{Hone} {et~al.}(2017){Hone}, {Kraus}, {Kreplin}, {Hofmann},
  {Weigelt}, {Harries}, \& {Kluska}}]{2017A&A...607A..17H}
{Hone}, E., {Kraus}, S., {Kreplin}, A., {et~al.} 2017, \aap, 607, A17,
  \dodoi{10.1051/0004-6361/201731531}

\bibitem[{{Ireland} \& {Kraus}(2008)}]{2008ApJ...678L..59I}
{Ireland}, M.~J., \& {Kraus}, A.~L. 2008, \apjl, 678, L59,
  \dodoi{10.1086/588216}

\bibitem[{{Jenkins} \& {Peacock}(2011)}]{2011MNRAS.413.2895J}
{Jenkins}, C.~R., \& {Peacock}, J.~A. 2011, \mnras, 413, 2895,
  \dodoi{10.1111/j.1365-2966.2011.18361.x}

\bibitem[{{Kluska} {et~al.}(2020){Kluska}, {Berger}, {Malbet}, {Lazareff},
  {Benisty}, {Le Bouquin}, {Absil}, {Baron}, {Delboulb{\'e}}, {Duvert},
  {Isella}, {Jocou}, {Juhasz}, {Kraus}, {Lachaume}, {M{\'e}nard},
  {Millan-Gabet}, {Monnier}, {Moulin}, {Perraut}, {Rochat}, {Pinte}, {Soulez},
  {Tallon}, {Thi}, {Thi{\'e}baut}, {Traub}, \& {Zins}}]{2020A&A...636A.116K}
{Kluska}, J., {Berger}, J.~P., {Malbet}, F., {et~al.} 2020, \aap, 636, A116,
  \dodoi{10.1051/0004-6361/201833774}

\bibitem[{{Kraus} {et~al.}(2012){Kraus}, {Calvet}, {Hartmann}, {Hofmann},
  {Kreplin}, {Monnier}, \& {Weigelt}}]{2012ApJ...746L...2K}
{Kraus}, S., {Calvet}, N., {Hartmann}, L., {et~al.} 2012, \apjl, 746, L2,
  \dodoi{10.1088/2041-8205/746/1/L2}

\bibitem[{{Kurucz}(1991)}]{1991ppag.conf...27K}
{Kurucz}, R.~L. 1991, in Precision Photometry: Astrophysics of the Galaxy, ed.
  A.~G.~D. {Philip}, A.~R. {Upgren}, \& K.~A. {Janes}, 27

\bibitem[{{Lazareff} {et~al.}(2017){Lazareff}, {Berger}, {Kluska}, {Le
  Bouquin}, {Benisty}, {Malbet}, {Koen}, {Pinte}, {Thi}, {Absil}, {Baron},
  {Delboulb{\'e}}, {Duvert}, {Isella}, {Jocou}, {Juhasz}, {Kraus}, {Lachaume},
  {M{\'e}nard}, {Millan-Gabet}, {Monnier}, {Moulin}, {Perraut}, {Rochat},
  {Soulez}, {Tallon}, {Thi{\'e}baut}, {Traub}, \& {Zins}}]{2017A&A...599A..85L}
{Lazareff}, B., {Berger}, J.-P., {Kluska}, J., {et~al.} 2017, \aap, 599, A85,
  \dodoi{10.1051/0004-6361/201629305}

\bibitem[{{Leinert} {et~al.}(1997){Leinert}, {Richichi}, \&
  {Haas}}]{1997A&A...318..472L}
{Leinert}, C., {Richichi}, A., \& {Haas}, M. 1997, \aap, 318, 472

\bibitem[{{Leisenring} {et~al.}(2012){Leisenring}, {Skrutskie}, {Hinz},
  {Skemer}, {Bailey}, {Eisner}, {Garnavich}, {Hoffmann}, {Jones}, {Kenworthy},
  {Kuzmenko}, {Meyer}, {Nelson}, {Rodigas}, {Wilson}, \&
  {Vaitheeswaran}}]{2012SPIE.8446E..4FL}
{Leisenring}, J.~M., {Skrutskie}, M.~F., {Hinz}, P.~M., {et~al.} 2012, in
  Society of Photo-Optical Instrumentation Engineers (SPIE) Conference Series,
  Vol. 8446, Society of Photo-Optical Instrumentation Engineers (SPIE)
  Conference Series, 4--19, \dodoi{10.1117/12.924814}

\bibitem[{{Maire} {et~al.}(2015){Maire}, {Skemer}, {Hinz}, {Desidera},
  {Esposito}, {Gratton}, {Marzari}, {Skrutskie}, {Biller}, {Defr{\`e}re},
  {Bailey}, {Leisenring}, {Apai}, {Bonnefoy}, {Brandner}, {Buenzli}, {Claudi},
  {Close}, {Crepp}, {De Rosa}, {Eisner}, {Fortney}, {Henning}, {Hofmann},
  {Kopytova}, {Males}, {Mesa}, {Morzinski}, {Oza}, {Patience}, {Pinna},
  {Rajan}, {Schertl}, {Schlieder}, {Su}, {Vaz}, {Ward-Duong}, {Weigelt}, \&
  {Woodward}}]{2015A&A...576A.133M}
{Maire}, A.-L., {Skemer}, A.~J., {Hinz}, P.~M., {et~al.} 2015, \aap, 576, A133,
  \dodoi{10.1051/0004-6361/201425185}

\bibitem[{{Malbet} {et~al.}(2007){Malbet}, {Benisty}, {de Wit}, {Kraus},
  {Meilland}, {Millour}, {Tatulli}, {Berger}, {Chesneau}, {Hofmann}, {Isella},
  {Natta}, {Petrov}, {Preibisch}, {Stee}, {Testi}, {Weigelt}, {Antonelli},
  {Beckmann}, {Bresson}, {Chelli}, {Dugu{\'e}}, {Duvert}, {Gennari},
  {Gl{\"u}ck}, {Kern}, {Lagarde}, {Le Coarer}, {Lisi}, {Perraut}, {Puget},
  {Rantakyr{\"o}}, {Robbe-Dubois}, {Roussel}, {Zins}, {Accardo}, {Acke},
  {Agabi}, {Altariba}, {Arezki}, {Aristidi}, {Baffa}, {Behrend}, {Bl{\"o}cker},
  {Bonhomme}, {Busoni}, {Cassaing}, {Clausse}, {Colin}, {Connot},
  {Delboulb{\'e}}, {Domiciano de Souza}, {Driebe}, {Feautrier}, {Ferruzzi},
  {Forveille}, {Fossat}, {Foy}, {Fraix-Burnet}, {Gallardo}, {Giani}, {Gil},
  {Glentzlin}, {Heiden}, {Heininger}, {Hernandez Utrera}, {Kamm}, {Kiekebusch},
  {Le Contel}, {Le Contel}, {Lesourd}, {Lopez}, {Lopez}, {Magnard}, {Marconi},
  {Mars}, {Martinot-Lagarde}, {Mathias}, {M{\`e}ge}, {Monin}, {Mouillet},
  {Mourard}, {Nussbaum}, {Ohnaka}, {Pacheco}, {Perrier}, {Rabbia}, {Rebattu},
  {Reynaud}, {Richichi}, {Robini}, {Sacchettini}, {Schertl}, {Sch{\"o}ller},
  {Solscheid}, {Spang}, {Stefanini}, {Tallon}, {Tallon-Bosc}, {Tasso},
  {Vakili}, {von der L{\"u}he}, {Valtier}, {Vannier}, \&
  {Ventura}}]{2007A&A...464...43M}
{Malbet}, F., {Benisty}, M., {de Wit}, W.-J., {et~al.} 2007, \aap, 464, 43,
  \dodoi{10.1051/0004-6361:20053924}

\bibitem[{{McGregor} {et~al.}(1984){McGregor}, {Persson}, \&
  {Cohen}}]{1984ApJ...286..609M}
{McGregor}, P.~J., {Persson}, S.~E., \& {Cohen}, J.~G. 1984, \apj, 286, 609,
  \dodoi{10.1086/162636}

\bibitem[{{Meeus} {et~al.}(2001){Meeus}, {Waters}, {Bouwman}, {van den Ancker},
  {Waelkens}, \& {Malfait}}]{2001A&A...365..476M}
{Meeus}, G., {Waters}, L.~B.~F.~M., {Bouwman}, J., {et~al.} 2001, \aap, 365,
  476, \dodoi{10.1051/0004-6361:20000144}

\bibitem[{{Millan-Gabet} {et~al.}(2001){Millan-Gabet}, {Schloerb}, \&
  {Traub}}]{2001ApJ...546..358M}
{Millan-Gabet}, R., {Schloerb}, F.~P., \& {Traub}, W.~A. 2001, \apj, 546, 358,
  \dodoi{10.1086/318239}

\bibitem[{{Monnier} {et~al.}(2009){Monnier}, {Tuthill}, {Ireland}, {Cohen},
  {Tannirkulam}, \& {Perrin}}]{2009ApJ...700..491M}
{Monnier}, J.~D., {Tuthill}, P.~G., {Ireland}, M., {et~al.} 2009, \apj, 700,
  491, \dodoi{10.1088/0004-637X/700/1/491}

\bibitem[{{Monnier} {et~al.}(2006){Monnier}, {Berger}, {Millan-Gabet}, {Traub},
  {Schloerb}, {Pedretti}, {Benisty}, {Carleton}, {Haguenauer}, {Kern},
  {Labeye}, {Lacasse}, {Malbet}, {Perraut}, {Pearlman}, \&
  {Zhao}}]{2006ApJ...647..444M}
{Monnier}, J.~D., {Berger}, J.-P., {Millan-Gabet}, R., {et~al.} 2006, \apj,
  647, 444, \dodoi{10.1086/505340}

\bibitem[{{Neyman} \& {Pearson}(1933)}]{1933RSPTA.231..289N}
{Neyman}, J., \& {Pearson}, E.~S. 1933, Philosophical Transactions of the Royal
  Society of London Series A, 231, 289, \dodoi{10.1098/rsta.1933.0009}

\bibitem[{{Ortiz-Le{\'o}n} {et~al.}(2017){Ortiz-Le{\'o}n}, {Dzib}, {Kounkel},
  {Loinard}, {Mioduszewski}, {Rodr{\'\i}guez}, {Torres}, {Pech}, {Rivera},
  {Hartmann}, {Boden}, {Evans}, {Brice{\~n}o}, {Tobin}, \&
  {Galli}}]{2017ApJ...834..143O}
{Ortiz-Le{\'o}n}, G.~N., {Dzib}, S.~A., {Kounkel}, M.~A., {et~al.} 2017, \apj,
  834, 143, \dodoi{10.3847/1538-4357/834/2/143}

\bibitem[{{Padgett} {et~al.}(1999){Padgett}, {Brandner}, {Stapelfeldt},
  {Strom}, {Terebey}, \& {Koerner}}]{1999AJ....117.1490P}
{Padgett}, D.~L., {Brandner}, W., {Stapelfeldt}, K.~R., {et~al.} 1999, \aj,
  117, 1490, \dodoi{10.1086/300781}

\bibitem[{{Pi{\'e}tu} {et~al.}(2003){Pi{\'e}tu}, {Dutrey}, \&
  {Kahane}}]{2003A&A...398..565P}
{Pi{\'e}tu}, V., {Dutrey}, A., \& {Kahane}, C. 2003, \aap, 398, 565,
  \dodoi{10.1051/0004-6361:20021551}

\bibitem[{{Ruane} {et~al.}(2017){Ruane}, {Mawet}, {Kastner}, {Meshkat},
  {Bottom}, {Femen{\'{\i}}a Castell{\'a}}, {Absil}, {Gomez Gonzalez}, {Huby},
  {Zhu}, {Jenson-Clem}, {Choquet}, \& {Serabyn}}]{2017AJ....154...73R}
{Ruane}, G., {Mawet}, D., {Kastner}, J., {et~al.} 2017, \aj, 154, 73,
  \dodoi{10.3847/1538-3881/aa7b81}

\bibitem[{{Sallum} \& {Eisner}(2017)}]{2017ApJS..233....9S}
{Sallum}, S., \& {Eisner}, J. 2017, The Astrophysical Journal Supplement
  Series, 233, 9, \dodoi{10.3847/1538-4365/aa90bb}

\bibitem[{{Sallum} {et~al.}(2017){Sallum}, {Eisner}, {Hinz}, {Sheehan},
  {Skemer}, {Tuthill}, \& {Young}}]{2017ApJ...844...22S}
{Sallum}, S., {Eisner}, J.~A., {Hinz}, P.~M., {et~al.} 2017, \apj, 844, 22,
  \dodoi{10.3847/1538-4357/aa7855}

\bibitem[{{Sallum} \& {Skemer}(2019)}]{2019JATIS...5a8001S}
{Sallum}, S., \& {Skemer}, A. 2019, Journal of Astronomical Telescopes,
  Instruments, and Systems, 5, 018001, \dodoi{10.1117/1.JATIS.5.1.018001}

\bibitem[{Sheehan(2018)}]{patrick_sheehan_2018_2455079}
Sheehan, P. 2018, {psheehan/pdspy: pdspy: A MCMC Tool for Continuum and
  Spectral Line Radiative Transfer Modeling}, \dodoi{10.5281/zenodo.2455079}

\bibitem[{{Smith} {et~al.}(2005){Smith}, {Balega}, {Duschl}, {Hofmann},
  {Lachaume}, {Preibisch}, {Schertl}, \& {Weigelt}}]{2005A&A...431..307S}
{Smith}, K.~W., {Balega}, Y.~Y., {Duschl}, W.~J., {et~al.} 2005, \aap, 431,
  307, \dodoi{10.1051/0004-6361:20041135}

\bibitem[{{Spalding} \& {Stone}(2019)}]{2019ascl.soft07008S}
{Spalding}, E., \& {Stone}, J. 2019, {Dewarp: Distortion removal and on-sky
  orientation solution for LBTI detectors}.
\newblock \doeprint{1907.008}

\bibitem[{{Stark} {et~al.}(2006){Stark}, {Whitney}, {Stassun}, \&
  {Wood}}]{2006ApJ...649..900S}
{Stark}, D.~P., {Whitney}, B.~A., {Stassun}, K., \& {Wood}, K. 2006, \apj, 649,
  900, \dodoi{10.1086/506926}

\bibitem[{{Strai{\v{z}}ys} {et~al.}(2003){Strai{\v{z}}ys}, {{\v{C}}ernis}, \&
  {Barta{\v{s}}i{\={u}}t{\.{e}}}}]{2003A&A...405..585S}
{Strai{\v{z}}ys}, V., {{\v{C}}ernis}, K., \& {Barta{\v{s}}i{\={u}}t{\.{e}}}, S.
  2003, \aap, 405, 585, \dodoi{10.1051/0004-6361:20030599}

\bibitem[{{Straizys} \& {Kuriliene}(1981)}]{1981Ap&SS..80..353S}
{Straizys}, V., \& {Kuriliene}, G. 1981, \apss, 80, 353,
  \dodoi{10.1007/BF00652936}

\bibitem[{{Terquem} \& {Bertout}(1993)}]{1993A&A...274..291T}
{Terquem}, C., \& {Bertout}, C. 1993, \aap, 274, 291

\bibitem[{{Thi{\'e}baut} \& {Young}(2017)}]{2017JOSAA..34..904T}
{Thi{\'e}baut}, {\'E}., \& {Young}, J. 2017, Journal of the Optical Society of
  America A, 34, 904, \dodoi{10.1364/JOSAA.34.000904}

\bibitem[{{Thompson} {et~al.}(1977){Thompson}, {Strittmatter}, {Erickson},
  {Witteborn}, \& {Strecker}}]{1977ApJ...218..170T}
{Thompson}, R.~I., {Strittmatter}, P.~A., {Erickson}, E.~F., {Witteborn},
  F.~C., \& {Strecker}, D.~W. 1977, \apj, 218, 170, \dodoi{10.1086/155668}

\bibitem[{{Trotta}(2008)}]{2008ConPh..49...71T}
{Trotta}, R. 2008, Contemporary Physics, 49, 71,
  \dodoi{10.1080/00107510802066753}

\bibitem[{{Tuthill} {et~al.}(2001){Tuthill}, {Monnier}, \&
  {Danchi}}]{2001Natur.409.1012T}
{Tuthill}, P.~G., {Monnier}, J.~D., \& {Danchi}, W.~C. 2001, \nat, 409, 1012,
  \dodoi{10.1038/35059014}

\bibitem[{{Tuthill} {et~al.}(2000){Tuthill}, {Monnier}, {Danchi}, {Wishnow}, \&
  {Haniff}}]{2000PASP..112..555T}
{Tuthill}, P.~G., {Monnier}, J.~D., {Danchi}, W.~C., {Wishnow}, E.~H., \&
  {Haniff}, C.~A. 2000, \pasp, 112, 555, \dodoi{10.1086/316550}

\bibitem[{{Ubeira-Gabellini} {et~al.}(2020){Ubeira-Gabellini}, {Christiaens},
  {Lodato}, {Ancker}, {Fedele}, {Manara}, \& {Price}}]{2020ApJ...890L...8U}
{Ubeira-Gabellini}, M.~G., {Christiaens}, V., {Lodato}, G., {et~al.} 2020,
  \apjl, 890, L8, \dodoi{10.3847/2041-8213/ab7019}

\bibitem[{{Vink} {et~al.}(2005){Vink}, {Drew}, {Harries}, {Oudmaijer}, \&
  {Unruh}}]{2005MNRAS.359.1049V}
{Vink}, J.~S., {Drew}, J.~E., {Harries}, T.~J., {Oudmaijer}, R.~D., \& {Unruh},
  Y. 2005, \mnras, 359, 1049, \dodoi{10.1111/j.1365-2966.2005.08969.x}

\bibitem[{{Vioque} {et~al.}(2018){Vioque}, {Oudmaijer}, {Baines},
  {Mendigut{\'\i}a}, \& {P{\'e}rez-Mart{\'\i}nez}}]{2018A&A...620A.128V}
{Vioque}, M., {Oudmaijer}, R.~D., {Baines}, D., {Mendigut{\'\i}a}, I., \&
  {P{\'e}rez-Mart{\'\i}nez}, R. 2018, \aap, 620, A128,
  \dodoi{10.1051/0004-6361/201832870}

\bibitem[{{Weigelt} {et~al.}(2011){Weigelt}, {Grinin}, {Groh}, {Hofmann},
  {Kraus}, {Miroshnichenko}, {Schertl}, {Tambovtseva}, {Benisty}, {Driebe},
  {Lagarde}, {Malbet}, {Meilland}, {Petrov}, \&
  {Tatulli}}]{2011A&A...527A.103W}
{Weigelt}, G., {Grinin}, V.~P., {Groh}, J.~H., {et~al.} 2011, \aap, 527, A103,
  \dodoi{10.1051/0004-6361/201015676}

\bibitem[{{Wolf} {et~al.}(2008){Wolf}, {Schegerer}, {Beuther}, {Padgett}, \&
  {Stapelfeldt}}]{2008ApJ...674L.101W}
{Wolf}, S., {Schegerer}, A., {Beuther}, H., {Padgett}, D.~L., \& {Stapelfeldt},
  K.~R. 2008, \apjl, 674, L101, \dodoi{10.1086/529188}

\end{thebibliography}

\appendix

\section{Data Reduction Steps}\label{app:red}
Here we describe the data reduction process in more detail, including updates to the LBT NRM pipeline since \citet{2017ApJS..233....9S}.

\subsection{Image Calibrations}

\subsubsection{Initial Corrections}
We first flat field the raw images using a flat constructed from the portions of science and calibrator frames that contain only sky background.
We perform bias, dark, and sky subtraction for each target pointing by subtracting the median of the top dither from the bottom dither, and vice versa.
We then apply a dewarping correction following the procedure described in \citet{2015A&A...576A.133M}, using \texttt{dewarp} \citep{2019ascl.soft07008S}.

\subsubsection{Additional Detector Systematics}
LMIRCam is an H2RG detector with 64-pixel wide channels having different analog-to-digital converters.
We measure and correct two types of systematic noise associated with these readout channels.
The first is pattern noise that repeats over each 64-pixel channel.
To characterize this, we separate the frame into 32 $64\times2048$ channels and take the median of each pixel across the 32 channels (excluding channels that contain images of the star).
We then subtract the median from all 32 readout channels.
We also correct for different bias levels in each of the readout channels.
We measure this by separating the frame into 32 $64\times2048$ channels, and calculating the median value of each channel (excluding rows that contain images of the star).
We then subtract the median from each channel in the image.

\subsubsection{Bad Pixels}
We lastly correct for bad pixels, replacing each pixel flagged as bad with the mean of the adjacent pixels.
We flag pixels as bad in two ways: (1) using a pre-generated bad pixel map, and (2) by examining each science frame individually.
We calculate the bad pixel map from the data we use to construct the sky flat.
From the distribution of all pixel values in the master flat, we flag all $3\sigma$ outliers as bad pixels.
We follow the same procedure with the master dark that was used to calibrate the sky flat, but using different $\sigma$ cuts on each side of the distribution ($2\sigma$ low, $3.5\sigma$ high) because of its skew.

For each science frame, we flag additional bad pixels after correcting those from the pre-generated map.
We compare each pixel to those in a $3\times3$ surrounding box.
We flag pixels that are greater than $2\sigma$ away from the mean of the box.
We choose a $3\times3$ box as the largest one that would not contain too much complex structure from the mask PSF, which would systematically increase $\sigma$ and make it more difficult to flag pixels.
We tested a variety of $\sigma$ cuts, and found that the correction did not change significantly for cuts of $2-3\sigma$ outliers.

\subsection{Choosing Fourier Sampling Coordinates}\label{sec:ftsamp}
We use the mask hole positions, diameters, and bandpasses to calculate synthetic power spectra for choosing Fourier sampling coordinates.
We create a realistic mask PSF by combining monochromatic mask PSFs across the bandpass. 
We take the $|\mathrm{FT}|^2$ of the mask PSF to calculate the power spectrum, and then find all pixels within $\sim50\%$ of the maximum for each baseline location.
We sample the observed complex visibilities at those pixels when generating squared visibilities and closure phases.

We compare a variety of synthetic power spectra to the mean observed power spectrum to check the sampling quality.
We allow for a non-zero mask rotation angle, and found that a small rotation ($\sim3.5^\circ$) provided the best match to the data.
However, even with a small rotation a mismatch existed for the inter-aperture baselines that could not be corrected by a simple scaling (e.g. bandpass adjustment).
Decreasing the horizontal separation (by $\sim0.5$ m) between holes on each of the two primaries corrected this mismatch. 
An effect like this could be caused by mask flexure, which would disproportionately affect the inter-aperture baselines.

\subsection{Squared Visibility and Closure Phase Generation}
We calculate the squared visibilities and closure phases by sampling the pixels described in Section \ref{sec:ftsamp}.
For the squared visibilities, we sum the power ($|\mathrm{FT}|^2$) for all pixels corresponding to each baseline.
We subtract a bias that we calculate by taking the mean of all power spectrum pixels without signal. 
We save these raw squared visibility amplitudes for each science frame, and also save normalized squared visibilities, which we calculate by dividing the zero-spacing power into the raw visibilities.

To calculate closure phases, for each triangle of baselines, we find all triangles of pixels whose ($u,v$) coordinates sum to (0,0).
We calculate a bispectrum for each pixel triangle by multiplying the FT values of the three pixels.
We then calculate an average bispectrum for each triangle and frame by averaging the bispectra of all the pixel triangles that close.
We describe our strategy for producing average squared visibilities and closure phases for each pointing in Section \ref{sec:avg}.

\subsection{Scan Averaging Strategy}\label{sec:avg}

Both the adaptive optics and co-phasing performance affect the relative quality of observables calculated for different frames and targets. 
AO performance affects the data quality for all baselines, impacting coherence for the intra-aperture fringes and the wavefront at the phase tracker. 
The phase tracking performance only influences the inter-aperture baselines.
Since the wavefront sensing and fringe tracking occur at different bandpasses (600-900 nm and 2.0-2.3 $\mu$m, respectively), science and reference PSF objects with different colors may have different relative V$^2$ and CP quality as a function of baseline length. 

\begin{figure}[ht]
\begin{center}
\begin{tabular}{c} 
\includegraphics[width=0.75\textwidth]{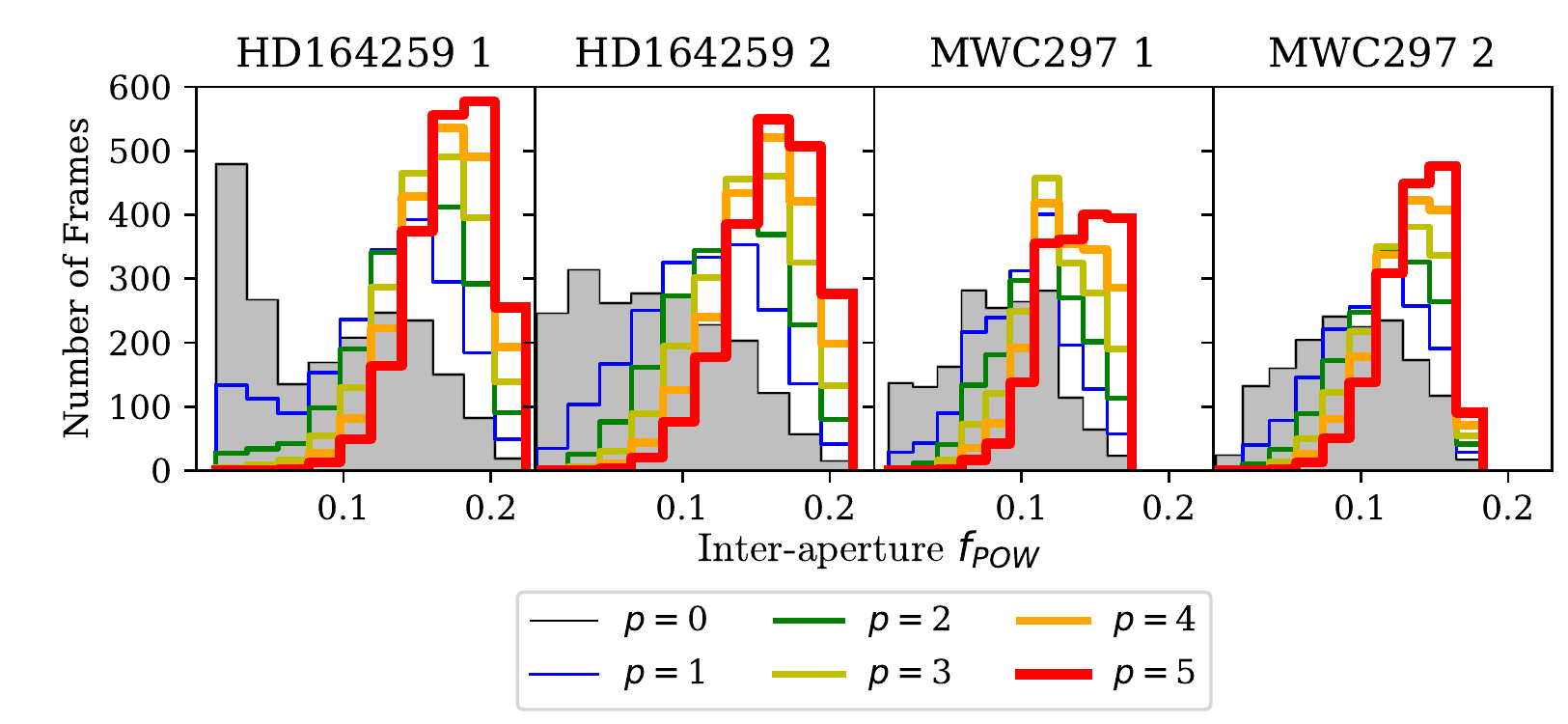}
\end{tabular}
\end{center}
\caption
{ \label{fig:pweight}
Weighted histograms showing the effective distributions of fractional inter-aperture power going into the average observables for each object pointing.
We use this fractional power to parameterize co-phasing performance, and to evaluate data averaging schemes intended to up-weight frames with better phase tracking.
Each line represents an averaging strategy that weights each frame by the total amplitudes on the inter-aperture baselines raised to the power $p$. The filled grey histogram shows the distribution without weighting ($p$ = 0), and the hollow histograms show increasing values of $p$ as the lines become redder and thicker.
}
\end{figure}

\begin{figure}[ht]
\begin{center}
\begin{tabular}{c} 
\includegraphics[width=0.7\textwidth]{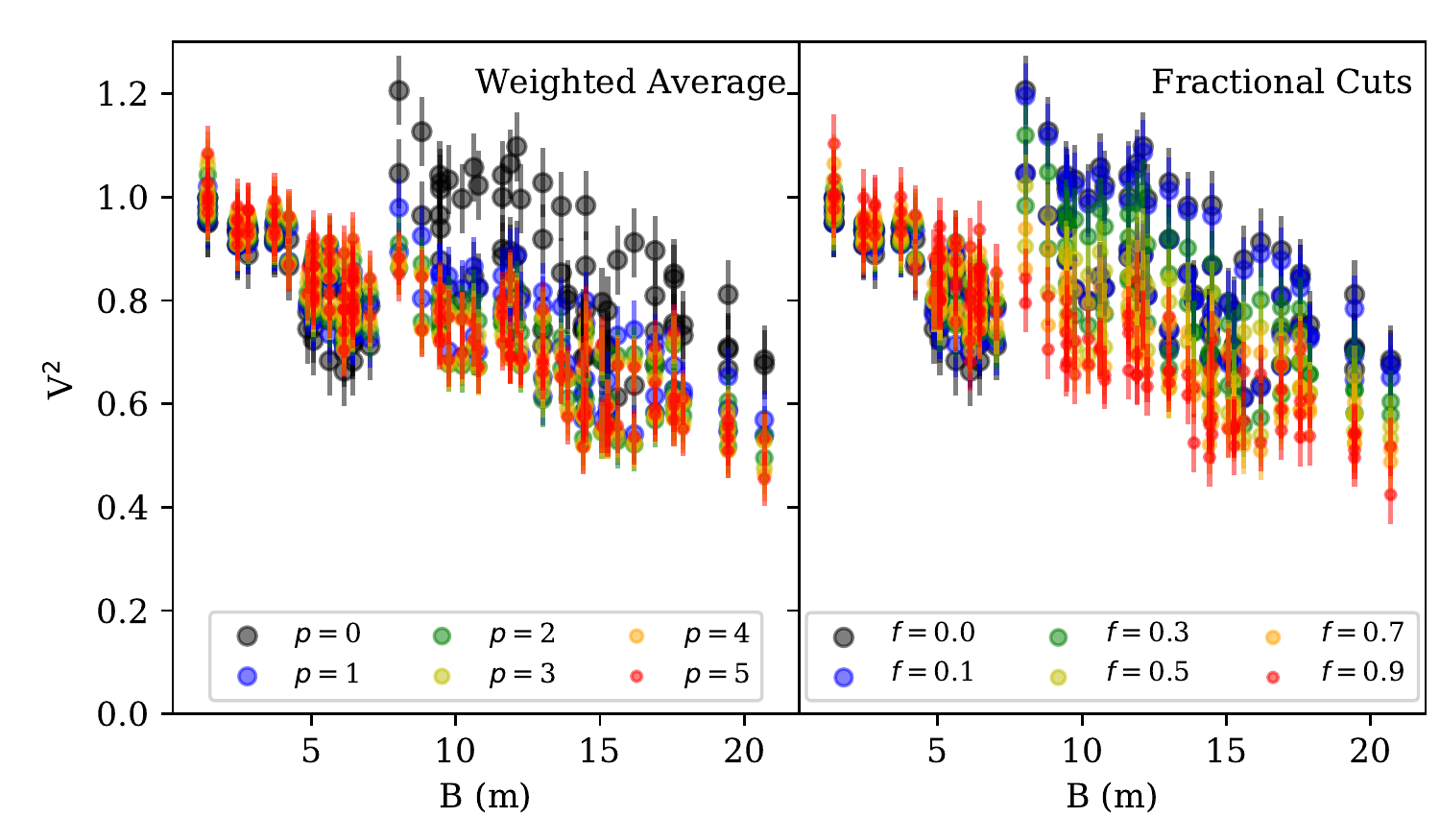}
\end{tabular}
\end{center}
\caption
{ \label{fig:v2comp}
Plotted points with error bars show the final squared visibilities for MWC 297 using the weighted averaging (left), and fractional cutting (right) approaches described in Section \ref{sec:avg}. In both panels, the largest grey points show the visibilities with no weighting scheme or cuts in the averaging process. Colors and sizes correspond to the power that the amplitudes are raised to in the weighting (left) and the fraction of lowest amplitude frames dropped (right). As the points become smaller and redder, the weighting and data cutting schemes become more aggressive. 
}
\end{figure}

While HD 164259 is significantly brighter than MWC 297 at $\lesssim1~\mu$m, its fainter Ks band flux leads to lower-quality long baseline observables than MWC 297.
We use the fraction of squared-visibility power that resides in the inter-aperture baselines ($f_{POW}$) as a representative measure of co-phasing performance for each frame. 
The worse co-phasing performance for HD 164259 can be seen in the larger number of low-$f_{POW}$ frames in the top-row shaded histograms in Figure \ref{fig:pweight}.
An averaging scheme that weights all individual frames equally would thus result in lower inter-aperture visibilities for HD 164259 just due to co-phasing performance. 
When these are divided into MWC 297 squared visibilities during calibration, the calibrated MWC 297 visibilities would appear higher (under-resolved) for inter-aperture baselines (Figure \ref{fig:v2comp}, grey points). 
This would also cause a systematic error in the inter-aperture closure phases, since they would be noisier for HD 164259, which would degrade any real CP signals in the calibrated MWC 297 data. 

To avoid these systematics in a more objective way than manual vetting, we apply a weighted averaging strategy aimed at evening the relative distributions of $f_{POW}$ for MWC 297 and HD 164259.
For the squared visibilities, we average the observables for the individual frames using the total visibility amplitudes in the inter-aperture baselines, raised to a power $p$, as a weight for each frame.
For the closure phases, we use the sum of all bispectrum amplitudes for triangles made up of two inter-aperture baselines, also raised to a power $p$.
We checked that this approach to the closure phase weighting is not significantly different than weighting both the squared visibilities and closure phases by simply the raw, inter-aperture visibility amplitudes.
Since closure phases are already weighted by bispectrum amplitudes during averaging, the improvement of this weighted scan averaging strategy was more pronounced for the squared visibilities than for the closure phases.

We use the raw sum of inter-aperture amplitudes (as opposed to the sum of inter-aperture amplitudes divided by the sum for all baselines) as the averaging weights, since the intra-aperture power for individual frames can vary as well (e.g. due to variability in AO performance).
Using the fractional inter-aperture power would thus unnecessarily upweight frames with equivalent inter-aperture power and lower intra-aperture power.
We test values of $p$ from 0 (no weighting) to 5 (aggressive up-weighting of coherent inter-aperture frames). 

\begin{figure}[ht]
\begin{center}
\begin{tabular}{c} 
\includegraphics[width=0.9\textwidth]{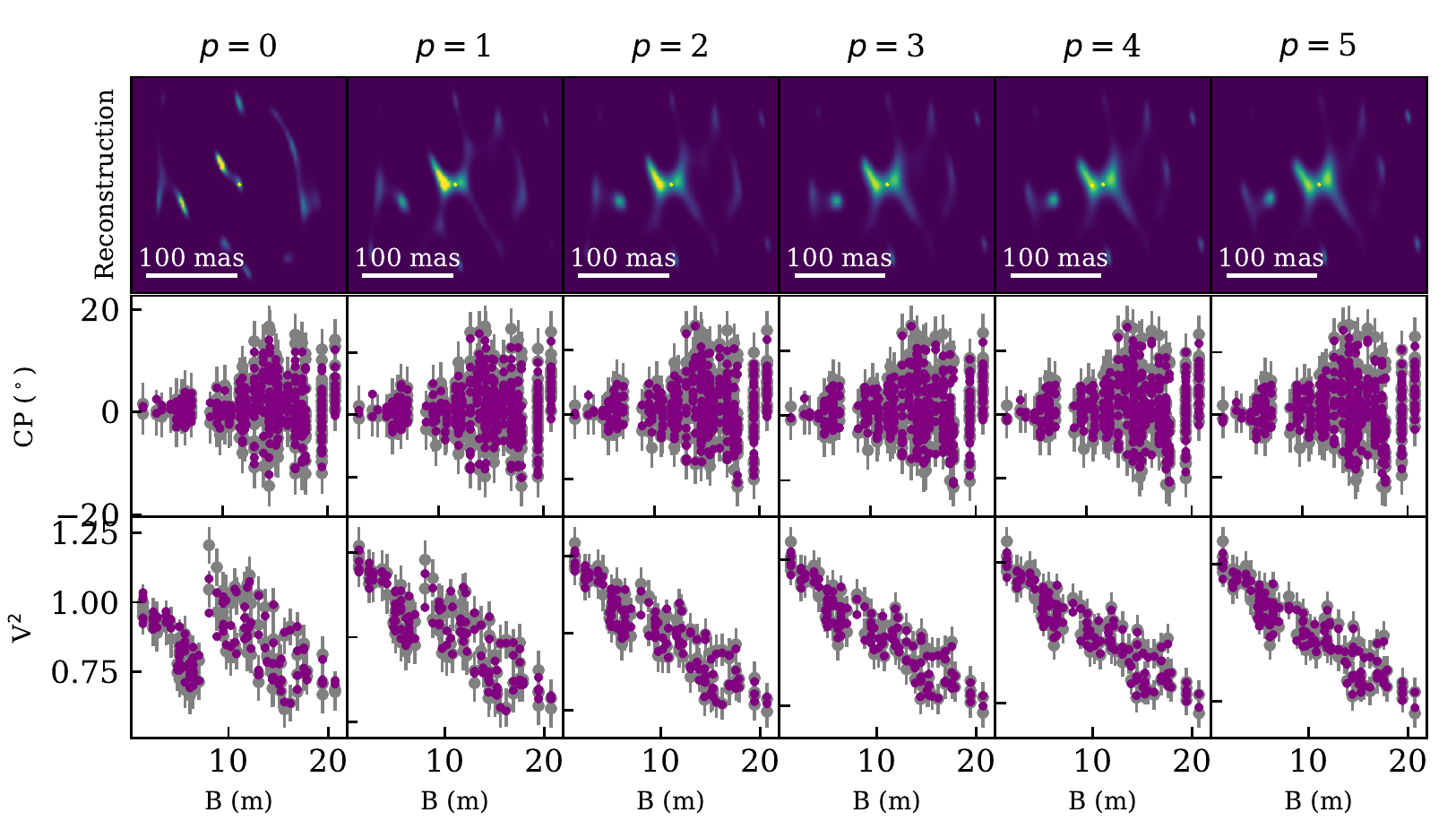}
\end{tabular}
\end{center}
\caption
{ \label{fig:precons}
BSMEM reconstructed images using the same prior image ($\delta$ + Circ. Gauss.~from Appendices \ref{app:modeling} and \ref{app:imrecon}) for scan averaging schemes with different values of $p$.
}
\end{figure}

Figure \ref{fig:pweight} shows the effective distributions of inter-aperture fractional power for each value of $p$.
We reconstructed images for all of these weighting schemes and found comparable results except for $p = 0$ (see Figure \ref{fig:precons}). 
Furthermore, we compared calibrated datasets for this scheme to a vetting scheme where we dropped some fraction of frames with the lowest inter-aperture amplitudes.
All values of $p\geq1$ were equivalent to dropping the lowest-quality $0.5-0.7$ of the frames.

Figure \ref{fig:v2comp} compares the final (calibrated) squared visibilities for these different methods and parameters.
We use the intermediate $p = 3$ weighted average as our representative dataset throughout the rest of the paper.
This value of $p$ leads to the lowest squared visibility errors estimated following the procedure in Appendix \ref{sec:errors}, with lower $p$ values suffering contamination from poorly co-phased frames, and higher $p$ values giving too much weight to a small number of frames.

\subsection{Calibration}
We calibrate both the squared visibilities and the closure phases using the ``polycal'' method described in \citet{2017ApJS..233....9S}.
We fit a polynomial to the calibrator observations as a function of time, and calculate its value at each target observation time.
We divide the resampled calibrator squared visibility into each science target squared visibility, and subtract each resampled calibrator closure phase from each science closure phase.
We note that one should generally correct for the angular size of the calibrator, and thus any resolved calibrator visibility signals, before carrying out the visibility calibration.
However, the calibrator angular size of 0.76 mas \citep{2004sf2a.conf..181D} corresponds to a squared visibility of 0.9994 on 23-m baselines.
We can thus treat the calibrator as an unresolved source.

Since only two calibrator pointings are available, we test $\mathrm{0^{th}}$- and $\mathrm{1^{st}}$-order polynomials, measuring the scatter in the calibrated data.
We use the polynomial order that minimizes the scatter in the calibrated MWC 297 observations, which for both the squared visibilities and the closure phases was the $\mathrm{0^{th}}$-order (constant) function.

\subsection{Error Bar Estimation}\label{sec:errors}
Systematic, rather than random noise sources dominate in NRM observations; the scatter in the final calibrated data is larger than the scatter across the cubes of images.
Assigning error bars based on the random variation in the observables would underestimate the error bars.
We thus use the distributions of calibrated closure phases and squared visibilities to estimate the errors on the data. 

First, to remove any mean signal, we calculate the mean squared visibility and closure phase for each baseline and triangle, respectively, by averaging the two pointings.
We then divide this out of the squared visibilities and subtract it from the closure phases.
We fit separate Gaussian functions to the resulting distributions of all squared visibilities and all closure phases separately, and assign each best-fit standard deviation as the error bar for all observables of its type (0.05 for squared visibilities, and $3.0^\circ$ for the closure phases).
Figure \ref{fig:finalobs} shows the final calibrated observations with their assigned error bars.

This simple approach may underestimate the error bars for the inter-aperture baselines and overestimate them for the intra-aperture baselines.
However, given the small number of observables we do not attempt to fit a more complex error model to the data.
We do, however, test the effects of both increasing and decreasing the errors as a function of baseline length, and find that this does not affect the results significantly for linear scalings with baseline length.
More aggressive error scalings, which do cause the reconstructed images to change, are unrealistic since they significantly upweight very small numbers of data points.

\section{Geometric Modeling}\label{app:modeling}
\subsection{Model Fitting}
We fit geometric models to the data to constrain MWC 297's morphology and to inform our image reconstruction tests. 
We use the results of previous long-baseline interferometric studies to inform our modeling assumptions.
IOTA observations utilizing 21-m and 38-m baselines at $1.65-2.2~\mu$m favored a Gaussian brightness distribution over a ring morphology for this object \citep{2001ApJ...546..358M}.
More recent VLTI datasets either showed no evidence for an inner clearing, or had a best-fit inner disk radius that would not be resolved by our observations \citep[$r_{in}\sim1-2$ mas;][]{2008A&A...485..209A,2011A&A...527A.103W,2017A&A...599A..85L}.
We thus explore simple $\delta$ function + Gaussian disk models (as opposed to $\delta$ + ring models).

We restrict the disk full-width at half-maximum (FWHM) to be less than $\sim70$ mas, slightly larger than the best-fit Gaussian size from $10.7~\mu$m Fizeau observations \citep{2009ApJ...700..491M} and larger than the dominant Gaussian components in fits to long-baseline interferometry data \citep[e.g.][]{2008A&A...485..209A}.  
While the $10.7~\mu$m observations suggested the presence of an extended halo (resolved by the shortest, $\sim1.8$m baselines in their array), it was only at the $\sim2\%$ level and thus we neglect it in the simple models presented here.
We also force the aspect ratio to be $>0.5$, since parametric fits to both the $10.7~\mu$m observations and long-baseline data are relatively axisymmetric \citep[$r > 0.77$; e.g.][]{2007A&A...464...43M,2020A&A...636A.116K}.

We apply a model consisting of a central delta function containing fractional flux $f_*$, and a skewed, Gaussian disk. 
The following equation defines the brightness distribution for the disk:
\begin{equation}
\begin{split}
I\left(x,y\right) = & \left(1+A_s \cos\left(\phi_s - \phi\right)\right)\\
& \times \exp\left[-\left(\left(\frac{x'}{\sqrt{2} \sigma_{x'}}\right)^2 + \left(\frac{y'}{\sqrt{2} r \sigma_{x'}}\right)^2\right)\right],
\end{split}
\end{equation}
where 
\begin{equation}
\begin{split}
x' = x \cos(\theta) - y \sin(\theta) \\
y' = x \sin(\theta) + y \cos(\theta),
\end{split}
\end{equation}
and where $(x,y)$ increase right and up in image space, respectively; $\theta$ is the position angle of the disk major axis, measured E of N; $\phi_s$ is the peak skew position angle measured E of N; $\phi = \arctan(y,x)$; $A_s$ is the skew amplitude, and $r$ is the minor to major axis ratio.
The full width at half maximum along the disk major axis is given by 
\begin{equation}
FWHM = 2\sqrt{\ln 2} \sigma_{x'}.
\end{equation}

We explore $\delta$ + Gaussian disk models of increasing complexity by first fixing $f_* = 0$,$r = 1$ and $A_s = 0$, and then relaxing constraints on the unresolved flux, axes ratio and skew.
We next allow for the presence of a companion-like feature in addition to the $\delta$ function and Gaussian disk, in the form of a second $\delta$ with a separation $s_c$, position angle (measured E of N) $\theta_c$, and contrast $c_c$ relative to the central $\delta$.
We restrict the companion separation to less than $\sim150$ mas since recent SPHERE observations detected no companions down to that inner working angle \citep{2020ApJ...890L...8U}.

Due to the large model parameter space, we use Markov-Chain Monte Carlo methods to explore possible models, rather than a grid search.
We use the open-source package \texttt{emcee} \citep{2013PASP..125..306F} in parallel-tempering mode, with 100 walkers and 10 temperatures for each model type.
This ensures that the space is well sampled even in the presence of local likelihood maxima.
We take the 16\% and 84\% contours in the T=1 chain as the 1$\sigma$ allowed range of model parameters.
For each model, we also calculate reduced $\chi^2$ to assess goodness of fit.

We use both likelihood ratios and Bayesian evidence ratios to estimate the likelihood that the model is preferred.
The performance of two nested models can be compared by calculating the ratio of the likelihood values for the two models, which in this case is the same as the difference in best-fit $\chi^2$ values \citep[e.g.][]{1933RSPTA.231..289N}.
This statistic is approximately $\chi^2$ distributed with $\Delta_{DOF}$ degrees of freedom, where $\Delta_{DOF}$ is the difference in degrees of freedom between the two models.
To calculate the likelihood ratio for any pair of models, we take the difference in log likelihood (the difference in $\chi^2$) between the best fits for the two models.
We calculate the significance level at which that model is preferred by comparing the $\Delta\chi^2$ to a distribution with $\Delta_{DOF}$ degrees of freedom.

Since log likelihood testing has a non-negative false positive probability \citep[e.g.][]{2011MNRAS.413.2895J}, we also calculate Bayesian evidence values for each model \citep[e.g][]{2008ConPh..49...71T}.
Bayesian evidence is the marginalized likelihood over the parameter space of interest.
For two different models, the odds of one model being preferred over another is proportional to the ratio of their evidence values.
We use \texttt{emcee} to estimate the Bayesian evidence from the various temperature chains using thermodynamic integration \citep[e.g.][]{2004AIPC..707...59G}.
We compare evidence values for the different models by examining differences in log evidence ($\log{Z}$). 

\subsection{Results}\label{sec:modelres}

\begin{deluxetable}{lccccccccc}
\tabletypesize{\footnotesize}
\tablecaption{Geometric Fit Results \label{tab:fpars}}
\tablehead{
\colhead{Model Type} &  \colhead{$FWHM$} & \colhead{$r$} & \colhead{$\theta$} & \colhead {$f_*$} & \colhead{$A_s$} & \colhead{$\phi_s$} & \colhead{$s_c$} & \colhead{$\theta_c$} & \colhead{$c_c$}  \\
\colhead{} & \colhead{(mas)} & \colhead{} & \colhead{($^\circ$)} & \colhead{} & \colhead{} & \colhead{($^\circ$)} & \colhead{(mas)} & \colhead{($^\circ$)} & \colhead{($\%$)} 
}
\startdata
Circ. Gauss. & $13.3\pm^{0.2}_{0.1}$ & -- & -- & -- & -- & -- & -- & -- & --   \\
$\delta$ + Circ. Gauss. & $43.2\pm^{2.5}_{2.0}$ & -- & -- & $0.774\pm^{0.008}_{0.009}$ & -- & -- & -- & -- & --   \\
$\delta$ + Non-Circ. Gauss. & $45.8\pm^{2.5}_{2.4}$ & $0.87\pm^{0.10}_{0.08}$ & $57\pm^{21}_{19} $ & $0.772\pm^{0.008}_{0.010}$  & -- & -- & -- & -- & --   \\
$\delta$ + Skew Gauss. & $44.7\pm^{2.5}_{2.4}$ & $0.88\pm^{0.05}_{0.07}$ & $57\pm^{21}_{24}$ & $0.767\pm^{0.009}_{0.010}$ & $0.23\pm^{0.04}_{0.05}$ & $-49\pm_{11}^{15}$ & -- & -- & --\\
$\delta$ + Skew Gauss. + Comp. & $38.3\pm^{3.4}_{2.9}$& $0.83\pm^{0.09}_{0.07}$ & $56\pm^{13}_{16}$ & $0.74\pm{0.01}$ & $0.25\pm^{0.05}_{0.03}$& $-61\pm_{11}^{19}$ & $58.5\pm^{2.4}_{1.7}$  & $101\pm2$ & $2.0\pm0.3$ \\
\enddata
\end{deluxetable}

\begin{deluxetable}{lcccccccc}
\tabletypesize{\footnotesize}
\tablecaption{Model Selection \label{tab:msel}}
\tablehead{
\colhead{Model Type} & \colhead{$dof$} &  \colhead{$\chi^2_{min}$} & \colhead{$\chi^2_r$} & \colhead{$\chi^2_{r,CP}$} & \colhead{$\chi^2_{r,V^2}$} &  \colhead{$\Delta\chi^2$\tablenotemark{a}} & \colhead{Sig.\tablenotemark{b}} & \colhead{$\log{Z}$} \\
}
\startdata
Circ. Gauss. & 571 & 1361.9 & 2.39 & 2.20 & 3.02 & -- & -- & -5134 \\
$\delta$ + Circ. Gauss. & 570 & 1169.0 & 2.05 & 2.21 & 1.54 & 192.9 & $>5\sigma$ & $-1792$  \\
$\delta$ + Non-circ. Gauss.  & 568 & 1165.5 & 2.05 & 2.22 & 1.53 & 3.5 & $>1\sigma$ & $-1175$ \\
$\delta$ + Skew Gauss. &  566 & 1132.9 & 2.00 & 2.16 & 1.57 &  32.6 & $>4\sigma$ & $-952$ \\
$\delta$ + Skew Gauss. + Comp &  563 & 1040.6 & 1.85 & 2.08 & 1.17 & 92.3 &  $>5\sigma$  & $-779$ \\
\enddata
\tablenotetext{a}{$\Delta\chi^2$ values list the decrease in minimum $\chi^2$ for a particular model compared to the model in the row above it.}
\tablenotetext{b}{Significance values for a particular model being preferred (based on $\Delta\chi^2$) compared to the model in the row above it.}
\end{deluxetable}

\begin{figure}[ht]
\begin{center}
\begin{tabular}{c} 
\includegraphics[width=0.75\textwidth]{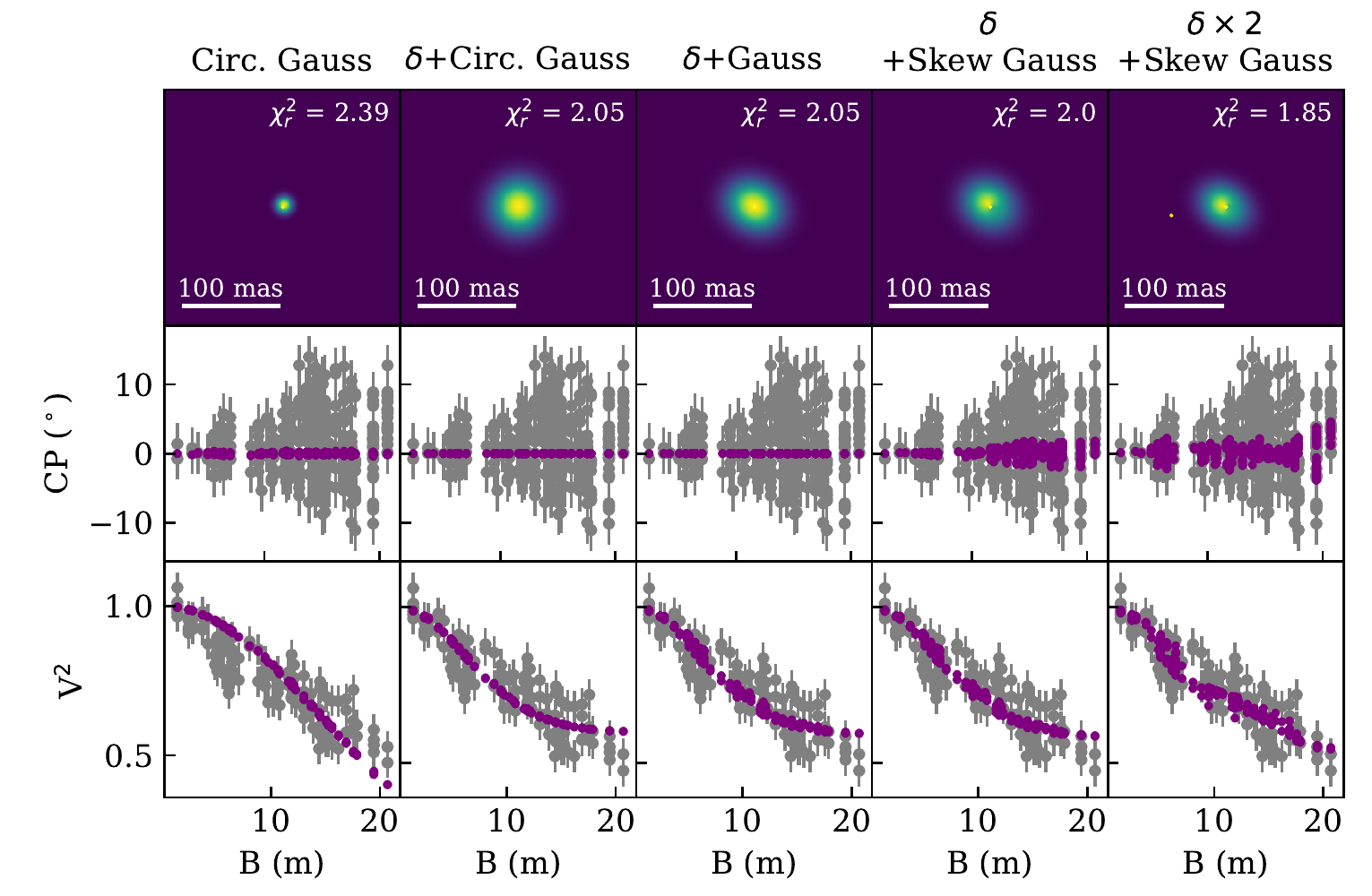}
\end{tabular}
\end{center}
\caption
{ \label{fig:dfits}
The top row shows the best-fit disk image for each of the models listed in Tables \ref{tab:fpars} and \ref{tab:msel}, scaled to bring out emission from the extended Gaussian components. The middle and bottom rows show the model closure phases and squared visibilities, respectively, (purple points) plotted over the observations (grey points with error bars). The best-fit reduced $\chi^2$ is listed in the top right corner of each image.
}
\end{figure}

Table \ref{tab:fpars} lists the best-fit parameters for the geometric model fitting whose corresponding images are shown in Figure \ref{fig:dfits}. Table \ref{tab:msel} lists model selection metrics: minimum $\chi^2$ values, reduced $\chi^2$ values (for all obserables, for just closure phases, and for just squared visibilities), $\Delta\chi^2$ values and their corresponding significance, and $\log{Z}$ values. 
The $\Delta\chi^2$ values in Table \ref{tab:msel} list the decrease in minimum $\chi^2$ associated with that model, compared to the model in the row above it.
The significance values show the significance with which one model is preferred compared to the simpler model above it, according to the likelihood ratio test. 

All of the models suggest the existence of a compact, unresolved component with a large fractional flux. 
The circular Gaussian model where $f_* = 0$ prefers a FWHM of $\sim13$ mas, placing most of the flux in the central region that is unresolved by the observations. 
This model is significantly worse than the models where $f_*$ is allowed to vary, which can be seen in the relatively high reduced $\chi^2$ and low $\log{Z}$ values.
This suggests the combination of an unresolved component like that seen in near-infrared long-baseline observations \citep[e.g.][]{2011A&A...527A.103W,2017A&A...599A..85L}, and an extended component similar to that seen at longer infrared wavelengths \citep[e.g.][]{2008A&A...485..209A,2009ApJ...700..491M}.

Indeed, all of the models with a central delta function point to an additional extended structure; their reduced $\chi^2$ and $\log{Z}$ values are much improved compared to the Circ. Gauss. model.
The majority of the flux ($\sim 0.74-0.77$ across the various models) is contained in the central unresolved component, consistent with the compactness of the Circ. Gauss. model.
The remaining fractional flux is distributed in an extended component with a characteristic size of $\sim40-45$ mas.
While there is a slight degeneracy between the $\delta$ function fractional flux ($f_*$) and the extended component FWHM, the two parameters are both relatively well constrained, varying together from $\textit{FWHM} = 39$ mas, $f_* = 0.76$ to $\textit{FWHM} = 48$ mas, $f_* = 0.79$.

The extended component must be relatively centro-symmetric to match the observations.
In particular, the squared visibilities would not show such consistent behavior with parallactic angle if the brightness distribution were highly asymmetric.
However, allowing for a non-circular Gaussian disk component leads to a large increase in $\log{Z}$ and a modest improvement in $\chi^2$. 
While the $\Delta\chi^2$ value only compares the best fits for two models, $\log{Z}$ compares the quality of the fit over the entire parameter space for two models.
This suggests that the two model sets have a similar quality best fit, but that a larger portion of the $\delta~+$ Non-circ.~Gauss. parameter space can match the observations.

The non-zero closure phases prefer models that allow for the extended emission to be skewed, with a $\sim25\%$ asymmetry along $\phi_s$. 
This is evidenced by both the large $\Delta\chi^2$ and the large increase in $\log{Z}$ between the first two models, whose disks do not have skew, and the last two models, whose disks allow for skew.
Furthermore, the $\delta~+$ Skew Gauss.~+ Comp.~model is strongly preferred by the data.
It has a $\Delta\chi^2$ of $\sim 92$ (corresponding a $>5\sigma$ significance with $\Delta_{DOF} = 3$), and a large increase in $\log{Z}$ compared to the $\delta~+$ Skew Gauss.~model.
The best-fit contrast for this companion relative to the central delta function component is $2.0\pm0.3\%$ or $4.25\pm0.15$ mag.

\begin{figure}[ht]
\begin{center}
\begin{tabular}{c} 
\includegraphics[width=0.45\textwidth]{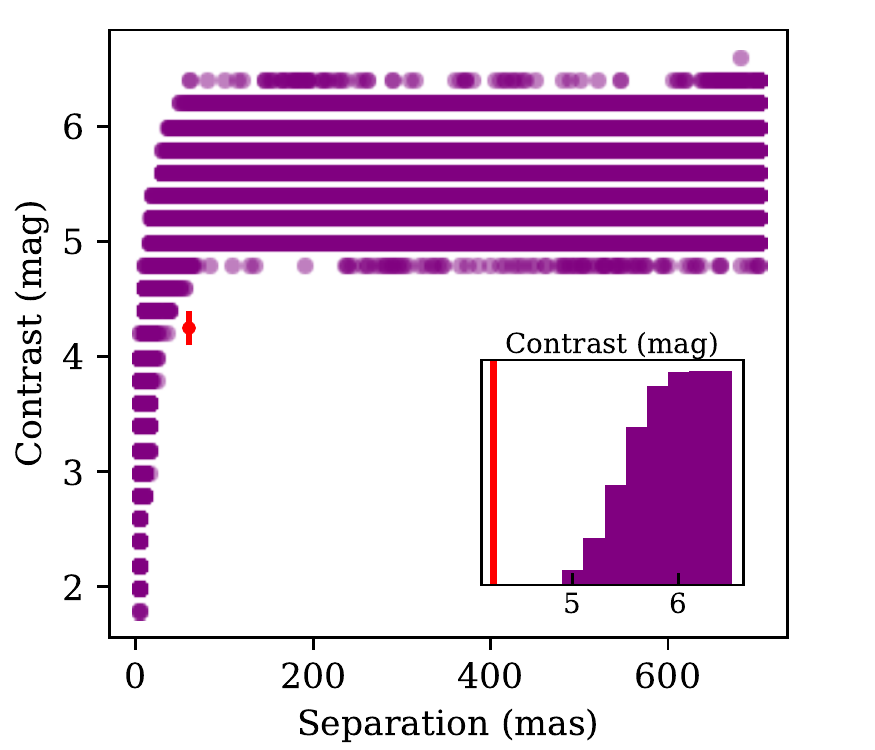}
\end{tabular}
\end{center}
\caption
{ \label{fig:fpp}
Main axis: Scattered purple points show the results of fitting companion models to Gaussian noise with the same level of scatter as the data. The red point with error bars shows the best fit separation and contrast for the companion signal. We calculate the false positive probability as the fraction of best fits that have the the same separation as the companion but lower contrast. Inset: The purple histogram shows the cumulative distribution of contrasts from fits to noise where the best fit had a separation equal to the companion candidate's. The solid red line shows the best fit contrast for the companion candidate.
}
\end{figure}

To test whether this companion signal could be caused by noise, we perform companion fits to Gaussian noise realizations.
We draw random observables from distributions of squared visibilities and closure phases with no mean signal and with the same level of scatter in the data ($\sigma_{V^2} = 0.05;~\sigma_{CP} = 3.0^\circ$).
We calculate the false positive probability as the fraction of best fits that have the the same separation as the companion but lower contrast.
Figure \ref{fig:fpp} shows the results; the companion contrast is 0.55 mag brighter than the brightest best fit to noise.
Because of the limited number of simulations, we can constrain the false positive rate to $<0.08\%$. 
We test whether noise plus a circumstellar disk model can reproduce the observed companion signal in Appendix \ref{app:simrtdisk}.

All of the best-fit models have large reduced $\chi^2$ values (1.85-2.05), suggesting that the data are under-fit, with a worse fit to the closure phases than to the squared visibilities.
MWC 297's underlying brightness distribution is likely more complex and asymmetric than a simple disk or even a disk plus companion. 
The models all under-fit the closure phases in particular, with $\chi^2_{r,CP}\sim2$ for all model types.
This points to more complex asymmetry than a disk plus companion model can represent.

\section{Additional Image Reconstruction Tests}\label{app:imrecon}

\subsection{Choice of Prior}\label{app:prior}

\begin{figure}[ht]
\begin{center}
\begin{tabular}{c} 
\includegraphics[width=0.8\textwidth]{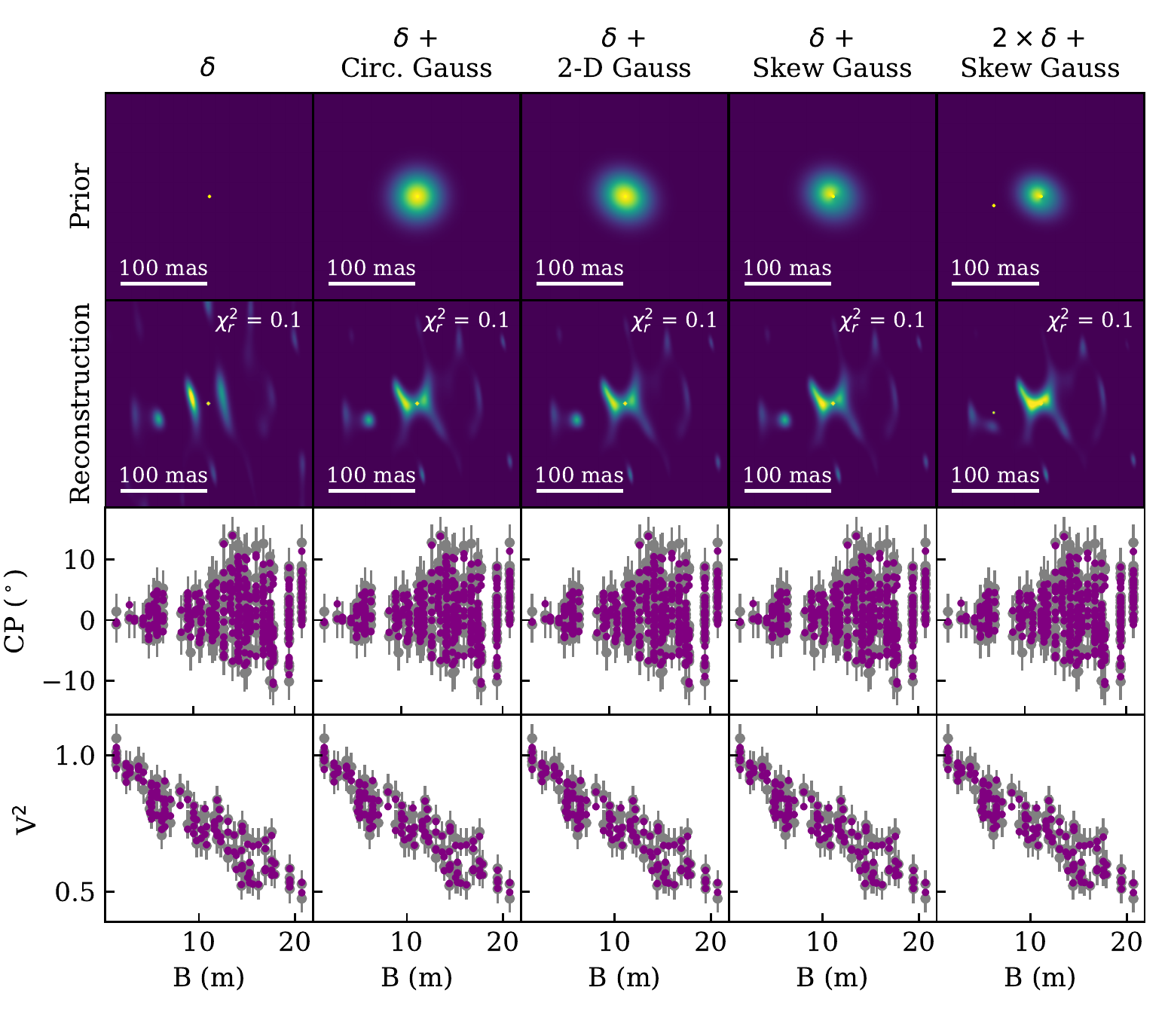}
\end{tabular}
\end{center}
\caption
{ \label{fig:pimages}
BSMEM reconstructed images for priors informed by geometric modeling. From top to bottom, each column shows for a single reconstruction, the prior image used, the resulting reconstructed image, the model closure phases (purple points) plotted against the data (grey points with error bars), and the model sqared visibilities (purple points) plotted against the data (grey points with error bars). All of the priors are best fits from models presented in Tables \ref{tab:fpars} and \ref{tab:msel}, with the exception of the first column, which is just a simple $\delta$ function. Each reconstructed image panel shows the reduced $\chi^2$ returned by BSMEM, which is defined as the $\chi^2$ of the reconstructed observables divided by the number of data points.
}
\end{figure}

To test the image fidelity, we reconstructed images with five additional priors other than the two-Gaussian prior whose results shown in Figure \ref{fig:finalim}: (1) a simple delta function, and (2-5) the best geometric fits from the models in Appendix \ref{app:modeling} that allow for a central $\delta$ function component.
Figure \ref{fig:pimages} presents the results.
In the $\delta$ image (Figure \ref{fig:pimages}, first column), the prior is aggressive enough to cause a flux deficit around the central component in the reconstruction.
However, the central, single pixel contains the same fractional flux as the central, beam-sized region in the other reconstructions.
Their less aggressive priors allow for flux close to the unresolved component.
Comparing the last two columns of Figure \ref{fig:pimages} illustrates this as well.
Introducing a delta function in the prior at the location of the companion feature leads it to become more compact in the reconstruction.

\begin{figure}[ht]
\begin{center}
\begin{tabular}{c} 
\includegraphics[width=0.8\textwidth]{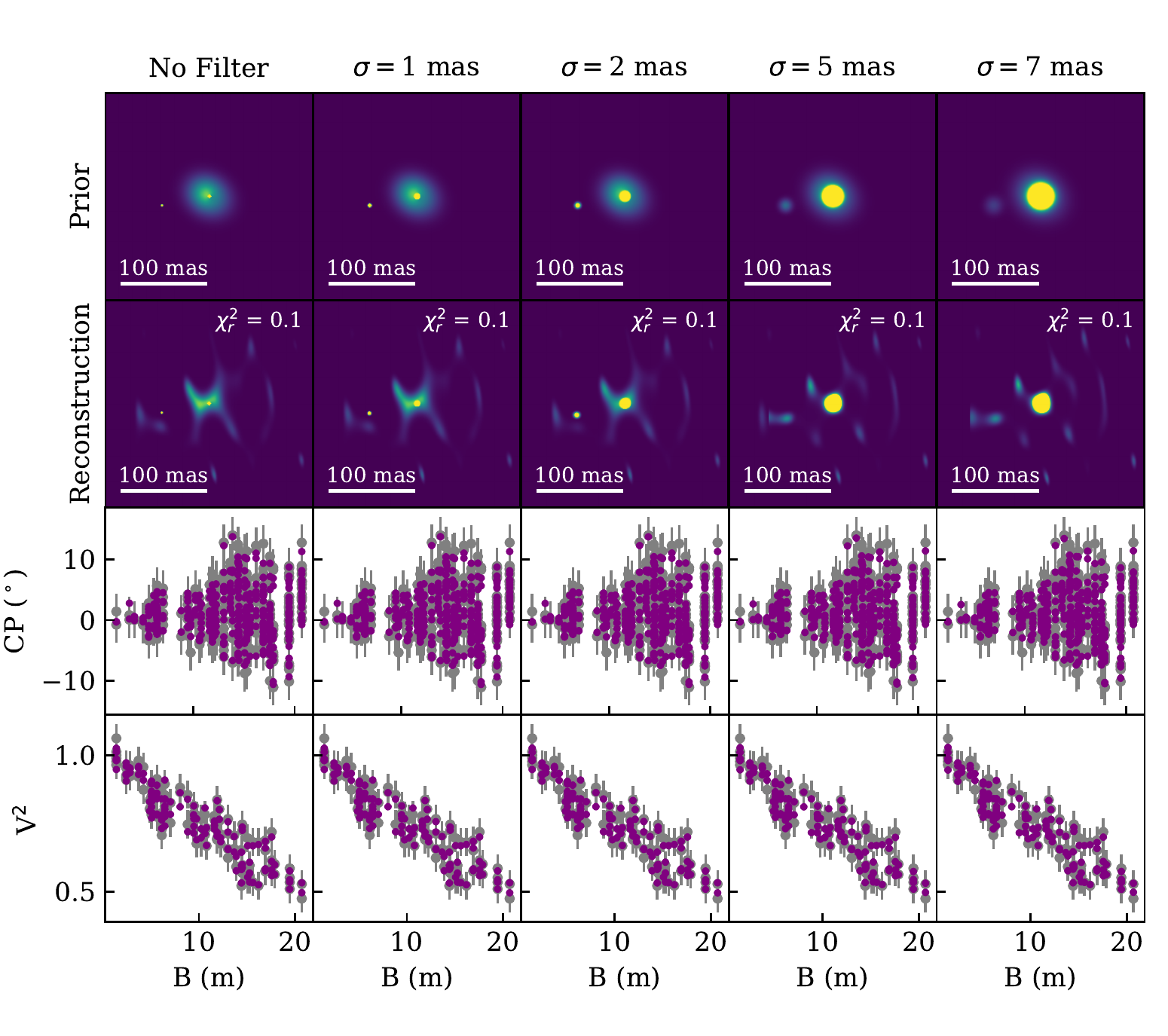}
\end{tabular}
\end{center}
\caption
{ \label{fig:gconvims}
BSMEM reconstructed images using the $2\times\delta$ + Skew Gauss prior convolved with Gaussian filters with a variety of $\sigma$ values. Each reconstructed image panel shows the reduced $\chi^2$ returned by BSMEM, which is defined as the $\chi^2$ of the reconstructed observables divided by the number of data points.
}
\end{figure}

To further test the influence of compactness in the prior, we convolve the $2\times\delta$ + Skew Gauss model with a Gaussian before reconstruction.
This leads to a prior image where the companion model feature and the central star both have $>1$ pixel extents.
Figure \ref{fig:gconvims} shows the results.
Indeed, as the two $\delta$ functions become more smeared out in the prior, BSMEM concentrates the flux less densely at that location in the resulting image. 
However, the total fractional flux in the central component and companion features does not change.

\subsection{Simulated Reconstructions of Geometric Models}\label{app:simrecon}

To further explore the impact of prior images, and to check whether simple geometric models could cause the structure in the reconstructions, we simulate image reconstructions for the models shown in Section \ref{sec:modelres}.
We sample each model with the same Fourier coverage and sky rotation as the observations.
We then add Gaussian noise to make the distributions of simulated observables match the data.
We reconstruct each model using each of the priors applied in Figure \ref{fig:pimages}.

\begin{figure}[ht]
\begin{center}
\begin{tabular}{c} 
\includegraphics[width=0.9\textwidth]{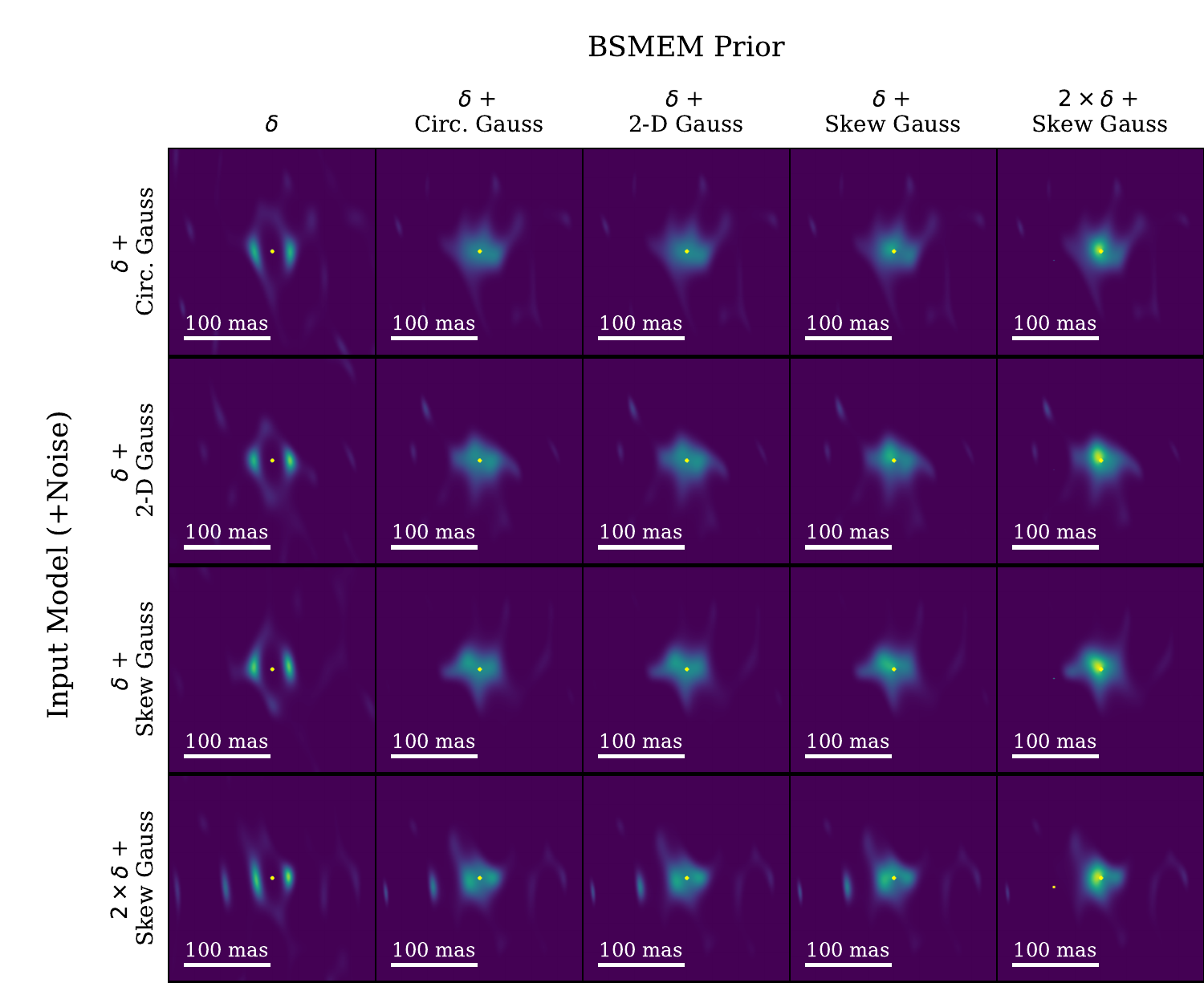}
\end{tabular}
\end{center}
\caption
{ \label{fig:dimages}
Reconstructed images from simulated observations of the best-fit disk models presented in Appendix \ref{sec:modelres}. We sampled the disk models with the same Fourier coverage and sky rotation as the data, and added enough noise to the simulated observables so that the distributions of closure phases and squared visibilities matched the observations. We then reconstructed images of each input model (rows) with all five priors used on the observations (columns). 
}
\end{figure}

Figure \ref{fig:dimages} shows the results of these tests. 
The $\delta$ function prior imposes a cleared region around the central, bright pixel for all models, while the less-aggressive priors do not. 
The prior image can introduce a small amount of skew when noise is present in observations of centrosymmetric models; the increased skew in the right end of the $\delta~+$ Circ. Gauss. row shows this. 
However, a skewed prior does not introduce as much skew for centrosymmetric objects as it does for truly asymmetric ones.
Comparing the images in the second column from the right shows this.

Including delta functions in a prior image always increases the flux in a single pixel at the same location in the reconstruction.
However, when no flux is present at that location in the true brightness distribution, the bright pixel is not significant compared to the noise in the rest of the image.
When the source does have significant emission at the prior delta function location, the single bright pixel in the reconstruction is significant. 
This can be seen in the rightmost column of Figure \ref{fig:dimages}. 
All four images contain one pixel with a brightness enhancement due to the prior, but it is only visible for the reconstruction of the disk plus companion model (fractional flux of $\sim0.019$).
The fractional fluxes in the same pixel for the other three images are $\sim0.00011-0.00018$, with the largest fractional flux in the reconstruction of the $\delta~+$ Skew Gauss.~model. 
In this case, BSMEM puts some of the asymmetric disk flux into a single bright pixel at the location of the $\delta$ function in the prior image.
However, the fractional flux in this pixel is still lower than the one in the $2\times\delta~+$ Skew Gauss.~reconstruction by a factor of $\sim100$.

\subsection{Simulated Reconstructions of Radiative Transfer Disk Models}\label{app:simrtdisk}
The simulations shown in Figure \ref{fig:dimages} show that simple geometric disk models cannot reproduce the data. 
Here we explore whether radiative transfer disk models, allowing for the presence of a disk rim, can match the observations without being inconsistent with previous VLTI datasets. 
We fit a coarse grid of radiative transfer models using the open-source software \texttt{RADMC-3D} \citep{2012ascl.soft02015D} and \texttt{pdspy} \citep{patrick_sheehan_2018_2455079}.
We then reconstruct images of the best-fit disk model by simulating observations with the same Fourier coverage as the data, and adding noise so that the simulated scatter matches the scatter in the data.

We explore two types of disk models: (1) a star + disk scenario with an inner radius allowed to vary, and (2) a star + inner disk from $\sim 0.2-2$ au + outer disk with an inner radius allowed to vary.
In all cases, we artificially increase the fractional flux of the star and/or inner disk, to account for the large compact fractional flux seen in the LBTI data and previous VLTI observations.
The best fit models from each of these categories have similar reduced $\chi^2$ values ($\sim2$), but only models from the second category are consistent with the VLTI closure phases from \citet{2020A&A...636A.116K}.

\begin{figure}[ht]
\begin{center}
\begin{tabular}{c} 
\includegraphics[width=0.9\textwidth]{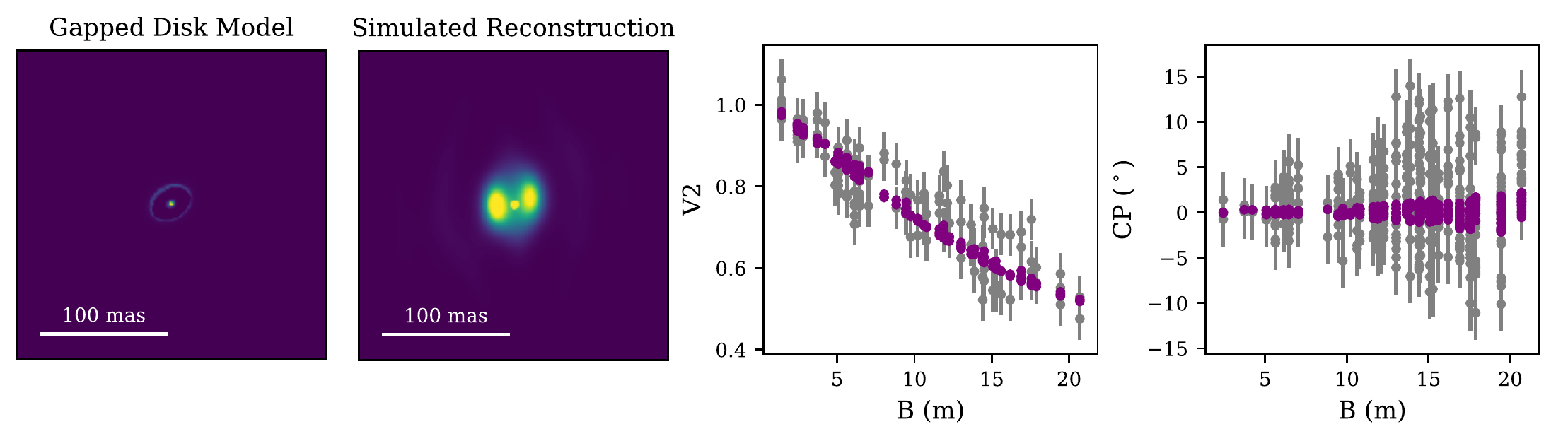}
\end{tabular}
\end{center}
\caption
{ \label{fig:RTfitim}
Left: the best-fit gapped disk radiative transfer model from the coarse model grid. Center left: the reconstructed image from simulated noiseless observations of the gapped disk model, shown on the same scale as the disk model. Center right: observed squared visibilities (grey points with error bars), and simulated squared visibilities for the disk model (purple points). Right: observed closure phases (grey points with error bars), and simulated closure phases for the disk model (purple points).
}
\end{figure}

We thus explore whether the best-fit gapped disk model could reproduce the observed LBTI reconstructed images.
Figure \ref{fig:RTfitim} shows the best-fit disk model, its model observables plotted against the data, the simulated image reconstruction from those observables.
The reduced $\chi^2$ for this model is $\chi^2_r = 1.93$.
This model has an outer disk inner radius of 8 au and an inclination of 40$^\circ$, and provides a relatively good match to the squared visibilities ($\chi^2_{r,V^2} = 1.28$), but not to the closure phases ($\chi^2_{r,CP} = 2.14$).
This suggests that the circumstellar disk has more complex structure than a simple gapped model. 
This is supported by the simulated image reconstruction, which does have a central depression, but does not show the same complex structure as the observed reconstructed image.

Injecting a 2\% companion at the location of the candidate in the data does not change this - it only introduces flux at the companion candidate location in the reconstruction.
A gapped disk plus companion model also still under-fits the data, with reduced $\chi^2$ values of: $\chi^2_r = 1.90;~\chi^2_{r,CP} = 2.12;~\chi^2_{r,V^2} = 1.21$.
Like the geometric models, the particularly high $\chi^2_{r,CP}$ suggests that the true circumstellar structure is more complex than a gapped disk plus companion.

Generating simulated reconstructions under conservative noise assumptions shows that a gapped disk model plus noise does not reliably reproduce the elongated structure in the observed reconstructed image.
We simulated a large number of observations of the gapped disk model, adding enough noise to match the distribution of squared visibilities and closure phases.
We note that assuming a best-fit model with $\chi^2_{r,CP}\sim2$ means that we must add a large amount of noise compared to the gapped disk signal (see Figure \ref{fig:RTfitim}) to match the distribution of closure phases.
For these simulated noise realizations, we estimate that less than $\sim10\%$ produced images with an asymmetric butterfly pattern similar to the observed reconstructed image. 
Reproducing the observations with this gapped disk scenario thus seems unlikely since it requires such a pathological noise realization.

We use these simulations to test whether a disk model plus noise could reproduce the companion signal in the observed reconstructed image. 
We reconstructed a large number of images from the best fit gapped disk model, with enough Gaussian noise added to match the scatter in the observed squared visibilities and closure phases. 
We then measured the fractional flux at the position of the companion candidate for each noise realization.
All of the observed fractional fluxes were lower than the $\sim2\%$ level measured in the companion candidate, demonstrating that a gapped disk model cannot reproduce the companion signal.
While this does not rule out extended circumstellar structure as the source of the companion candidate, it shows that reproducing the signal with circumstellar material requires a more complex morphology than a simple disk.

\subsection{L-curve Reconstructed Images}\label{app:lcurve}

\begin{figure}[ht]
\begin{center}
\begin{tabular}{c} 
\includegraphics[width=0.9\textwidth]{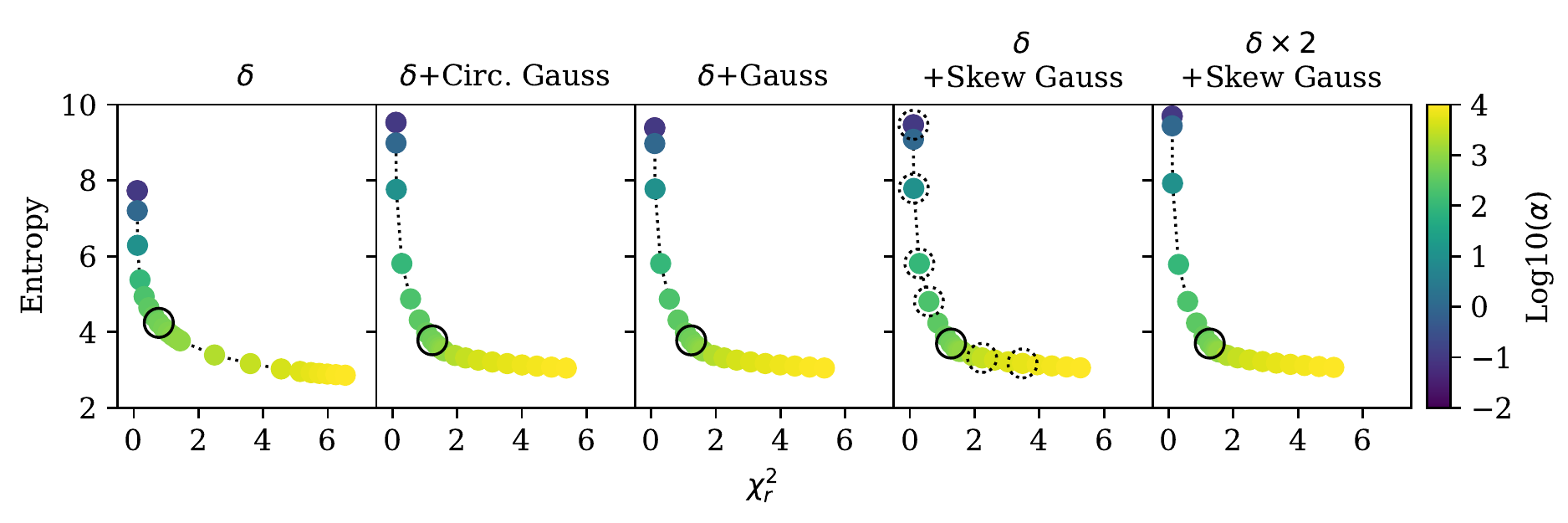}
\end{tabular}
\end{center}
\caption
{ \label{fig:Lcurves}
L-curves showing final image entropy versus reduced $\chi^2$ (the $\chi^2$ of the BSMEM-reconstructed observables divided by the number of data points) for reconstructions with different prior images (Appendix \ref{sec:modelres}). The solid-line hollow circle in each panel corresponds to the regularization value used in Figure \ref{fig:Lpimages}. The dotted-line hollow circles in the fourth panel show the regularization values explored in Figure \ref{fig:hyper}.
}
\end{figure}

In addition to BSMEM's automated hyperparameter optimization, for each prior image we also use the ``L-curve'' method to explore entropy hyperparameters \citep[e.g.][]{Hansen1992,2017JOSAA..34..904T}.
This involves plotting the image regularizer function (entropy) versus $\chi^2$, which has an L-shape.
The vertical portion of the L-curve, where $\chi^2$ values do not change with regularization, is dominated by the likelihood function and is under-regularized. 
In contrast, the horizontal section of the L-curve, with rapidly-changing $\chi^2$ values, is dominated by the regularization and is thus over-regularized.
We use the elbow in the L-curve, which balances the influence of the likelihood and regularizer, as the final reconstruction parameters.
For all prior images and using the error bars presented in Appendix \ref{sec:errors}, the elbow corresponds to an entropy hyperparameter of $\alpha\sim500$ (Figure \ref{fig:Lcurves}).

\begin{figure}[ht]
\begin{center}
\begin{tabular}{c} 
\includegraphics[width=0.75\textwidth]{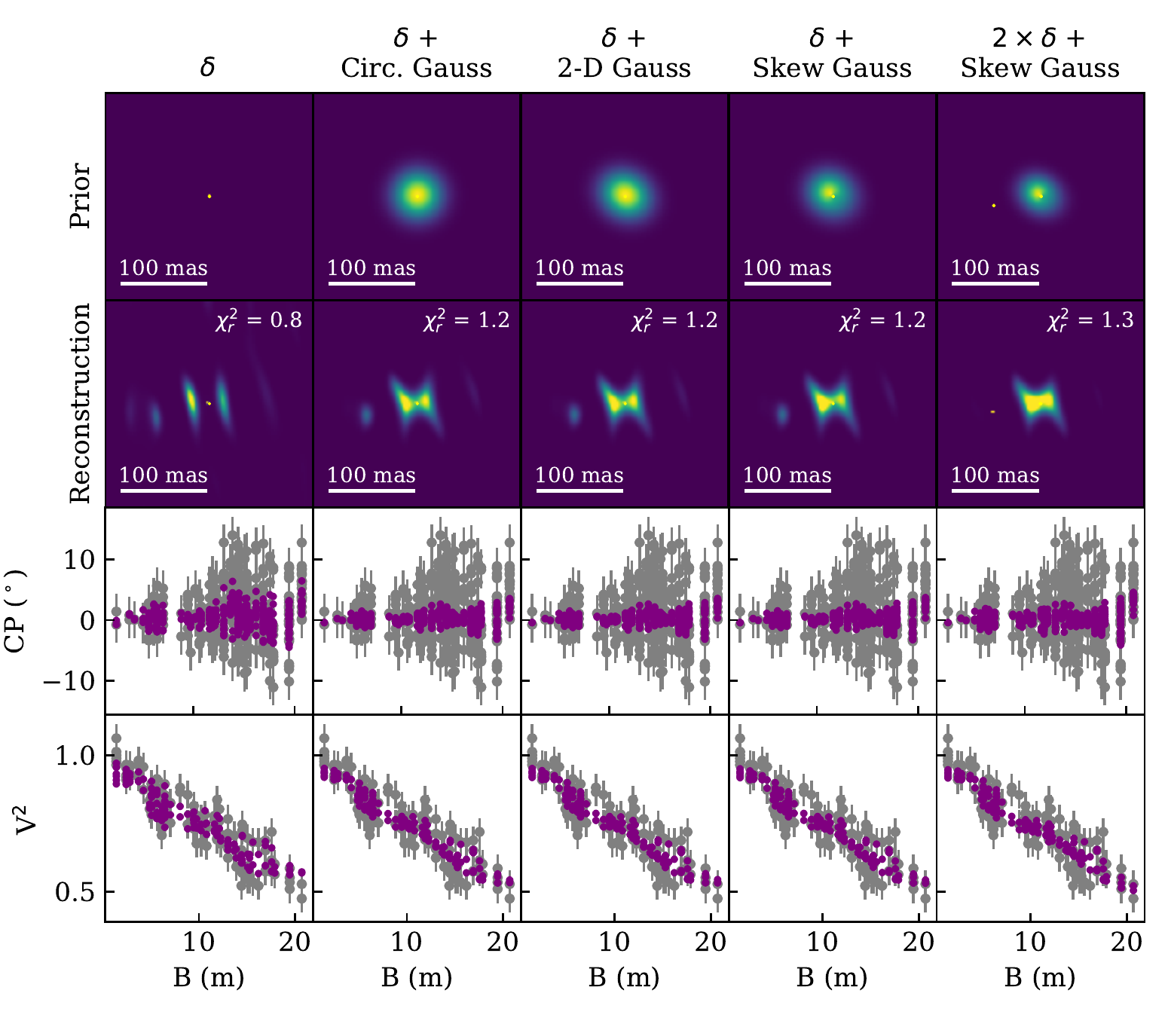}
\end{tabular}
\end{center}
\caption
{ \label{fig:Lpimages}
BSMEM reconstructed images using L-curve $\alpha$ optimization and priors informed by geometric modeling. From top to bottom, each column shows for a single reconstruction, the prior image used, the resulting reconstructed image, the model closure phases (purple points) plotted against the data (grey points with error bars), and the model squared visibilities (purple points) plotted against the data (grey points with error bars). All of the priors are best fits from models presented in Tables \ref{tab:fpars} and \ref{tab:msel}, with the exception of the first column, which is just a simple $\delta$ function. Each reconstructed image panel shows the reduced $\chi^2$ returned by BSMEM, which is defined as the $\chi^2$ of the reconstructed observables divided by the number of data points.
}
\end{figure}

Figure \ref{fig:Lpimages} shows the reconstructed images for $\alpha=500$ using the error bars estimated in Appendix \ref{sec:errors}, (which correspond roughly to reconstructed image reduced $\chi^2$ values of 1, where reduced $\chi^2$ is defined by BSMEM as the $\chi^2$ of the reconstructed observables divided by the number of data points).
The reconstructed image observables (purple scattered points in the lower two rows) have significantly lower scatter than both the observations and the reconstructed observables for BSMEM's automated hyperparameter optimization (Figure \ref{fig:finalimobs}).
This is because the automated optimization chooses hyperparameter values of $\sim10-15$, which lie in the under-regularized region of the L-curve.
However, the structure in the innermost $\sim250$ mas in these images is not significantly different from that shown in Figure \ref{fig:finalim}.

\subsection{Image Dependence on $\alpha$ and $\sigma$}\label{app:errs}

\begin{figure}[ht]
\begin{center}
\begin{tabular}{c} 
\includegraphics[width=\textwidth]{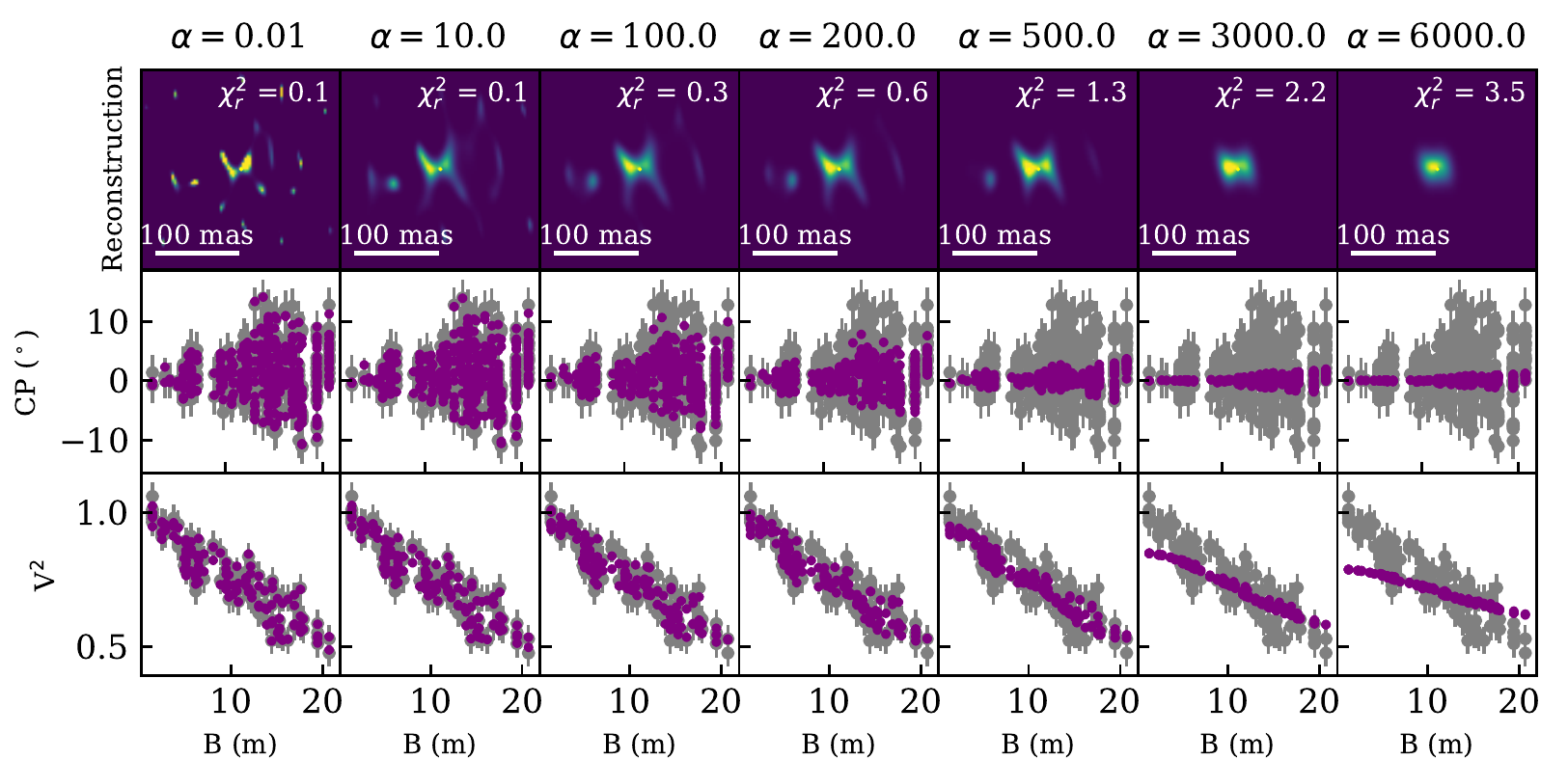}
\end{tabular}
\end{center}
\caption
{ \label{fig:hyper}
BSMEM reconstructed images using the $\delta$ + Skew Gauss prior with different entropy hyperparameters. The top row shows the reconstructed image, the middle row shows closure phases (purple points) plotted against the data (grey points with error bars), and the bottom row shows model squared visibilities (purple points) plotted against the data (grey points with error bars). Each reconstructed image panel shows the reduced $\chi^2$ returned by BSMEM, which is defined as the $\chi^2$ of the reconstructed observables divided by the number of data points.
}
\end{figure}

We test the effects of changing the entropy hyperparameter and the estimated error bars on the reconstructions (Figure \ref{fig:hyper}).
For smaller hyperparameters (where the reconstruction is more dominated by the likelihood function than the regularizer), the reconstructed image better matches the scatter in the data.
BSMEM accomplishes this by adding low levels of high-frequency signal throughout the field of view, which can create large closure phase signals.
Conversely, very large hyperparameters reduce the information in the reconstructed image to the point that it no longer qualitatively matches the observations.
Varying the entropy hyperparameter close to the elbow of the L-curve does not qualitatively change the reconstructed images.

\begin{figure}[ht]
\begin{center}
\begin{tabular}{c} 
\includegraphics[width=0.4\textwidth]{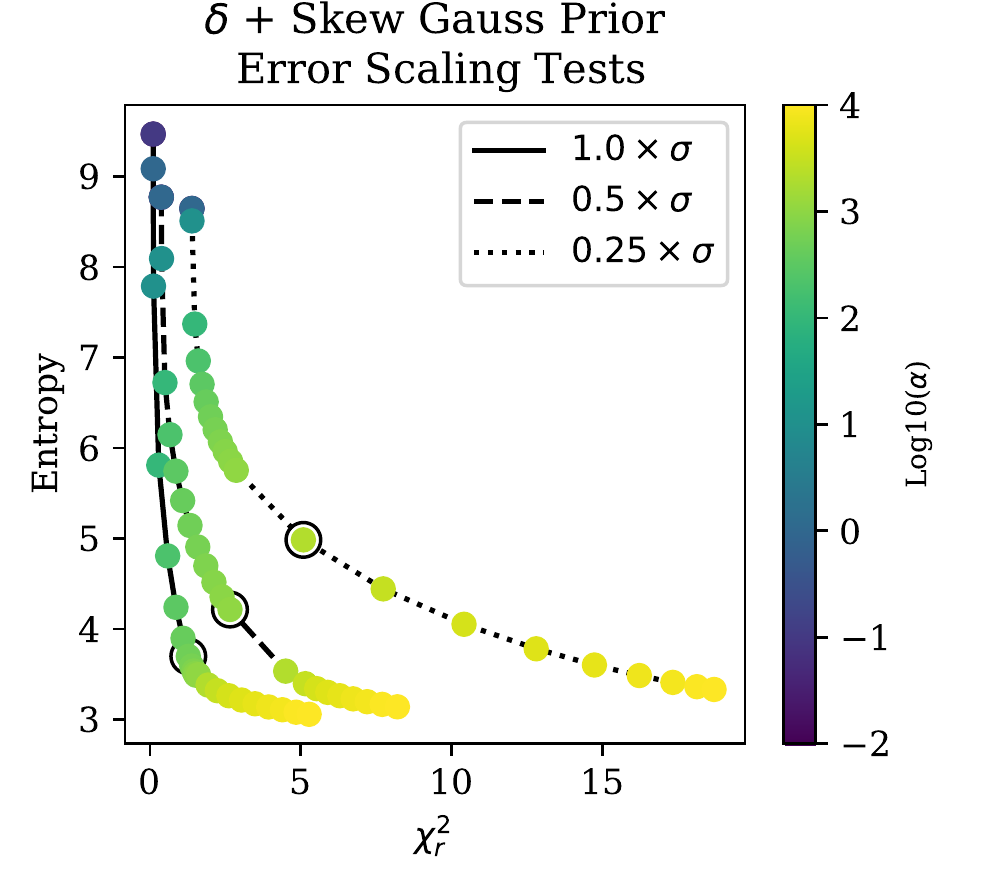}
\end{tabular}
\end{center}
\caption
{ \label{fig:Lcerrs}
L-curves for the reconstruction using the $\delta$ + Skew Gauss prior with different error bar scalings. The solid curve (leftmost) shows the results for error bars assigned according to Appendix \ref{sec:errors}, the dashed line (middle) shows the results for a $0.5\times$ scaling, and the dotted line (rightmost) shows the results for a $0.25\times$ scaling. The solid-line hollow circle shows the entropy hyperparameters at the L-curve elbows - 2000, 1000, and 500 for $0.25\times$, $0.5\times$ and $1.0\times$ scalings, respectively. These reconstructions are shown in Figure \ref{fig:imerrs}.
}
\end{figure}

\begin{figure}[ht]
\begin{center}
\begin{tabular}{c} 
\includegraphics[width=0.5\textwidth]{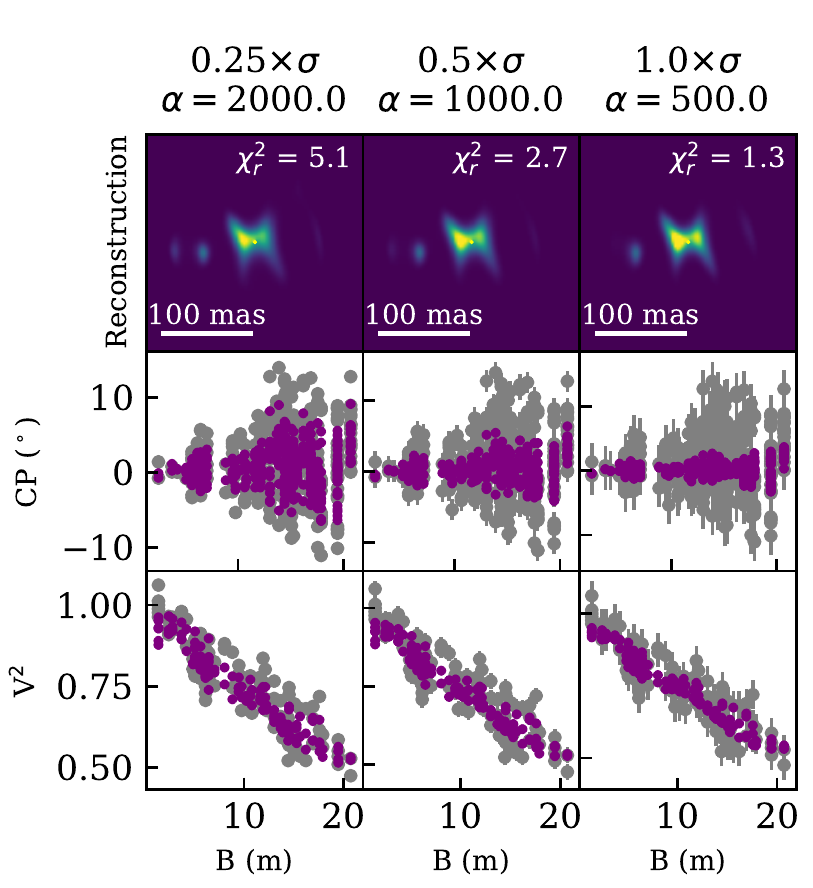}
\end{tabular}
\end{center}
\caption
{ \label{fig:imerrs}
BSMEM reconstructed images using a $\delta$ + Skew Gauss prior and the L-curve method for the error scalings shown in Figure \ref{fig:Lcerrs}. The top row shows the reconstructed image, the middle row shows closure phases (purple points) plotted against the data (grey points with error bars), and the bottom row shows model squared visibilities (purple points) plotted against the data (grey points with error bars). Each reconstructed image panel shows the reduced $\chi^2$ returned by BSMEM, which is defined as the $\chi^2$ of the reconstructed observables divided by the number of data points.
}
\end{figure}

Incorrect error bars change the location of the L-curve elbow and affect the quality of the reconstruction.
Underestimated error bars move the L-curve elbow toward higher regularization hyperparameters and larger reduced $\chi^2$ values.
We demonstrate this in Figure \ref{fig:Lcerrs}, which shows L-curves for the nominal error bars, and error bars that are a factor of two and four lower. 
For these reconstructions, BSMEM better matches the observations again by adding low levels of high-frequency signal to the image, without a significant qualitative change (Figure \ref{fig:imerrs}).

\end{document}